\documentclass[twocolumn,usenatbib]{mn2e}
\usepackage{subfigure}
\usepackage{color}
\usepackage{amsmath}
\usepackage{amssymb}
\usepackage{mathptmx}
\usepackage{times}
\usepackage{epsfig}
\usepackage{mathrsfs}
\usepackage{epstopdf}
\usepackage{txfonts}

\usepackage{xcolor}
\usepackage[colorlinks]{hyperref}
\hypersetup{ colorlinks, linkcolor=blue, citecolor=blue, anchorcolor=blue }

\usepackage{natbib}
\bibpunct{(}{)}{;}{a}{}{,}

\DeclareMathAlphabet{\mathsc}{OT1}{cmr}{m}{sc}
\def\testbx{bx}%
\DeclareRobustCommand{\ion}[2]{%
\relax\ifmmode
\ifx\testbx\f@series
{\mathbf{#1\,\mathsc{#2}}}\else
{\mathrm{#1\,\mathsc{#2}}}\fi
\else\textup{#1\,{\mdseries\textsc{#2}}}%
\fi}

\newcommand{\Nai}{\ion{Na}{i}}

\defcitealias{Liu15c}{LMS15}
\defcitealias{Rabinak12}{RLW12}
\defcitealias{Pan12}{PRT12}

\title[The progenitor of SN~2012cg]{Constraining the Progenitor of the Type Ia Supernova SN~2012cg}
\author[Z. W. Liu]{Zheng-Wei Liu\thanks{E-mail:zwliu@ynao.ac.cn} and Richard J. Stancliffe
                    \\
$^{1}$Argelander-Institut f\"ur Astronomie, Auf dem H\"ugel 71, D-53121, Bonn, Germany\\
}

\begin{document}



\maketitle

\label{firstpage}

\begin{abstract}

The nature of the progenitors of Type Ia supernovae (SNe Ia) is not yet fully understood. 
In the single-degenerate (SD) scenario, the collision of the SN ejecta with its companion star is 
expected to produce detectable ultraviolet (UV) emission in the first few days after the 
SN explosion within certain viewing angles. It was recently found that the $B-V$ colour of the nearby SN Ia SN~2012cg at about sixteen days 
before the maximum $B$-band brightness was about 0.2\,mag bluer than those of other normal SNe Ia, which was reported as the 
first evidence for excess blue light from the interaction of normal SN Ia ejecta with its companion star. In this work, we compare current observations for SN~2012cg 
from its pre-explosion phase to the late-time nebular phase with theoretical predictions from binary evolution and population synthesis 
calculations for a variety of popular progenitor scenarios. We find that a main-sequence donor or a carbon-oxygen white
dwarf donor binary system is more likely to be the progenitor of SN~2012cg. However, both scenarios also predict properties 
which are in contradiction to the observed features of this system. Taking both theoretical and observational uncertainties into account, 
we suggest that it might be too early to conclude that SN~2012cg was produced from an explosion of a Chandrasekhar-mass 
white dwarf in the SD scenario. Future observations and improved detailed theoretical modelling are still required to
place a more stringent constraint on the progenitor of SN~2012cg.

\end{abstract}


\section{INTRODUCTION}
 \label{sec:introduction}
Type Ia supernovae (SNe Ia) are important cosmological probes that led to the discovery of the accelerating 
expansion of the Universe \citep{Ries98, Schm98, Perl99}. However, 
the nature of SN~Ia progenitors and the physics of their explosion mechanisms are still a mystery \citep[e.g.][for a review]{Hill00, Maoz14}. 
There is a general consensus that SNe~Ia arise from thermonuclear explosions of white dwarfs (WDs) in 
binary systems \citep{Hoyl60, Nomo82}. 
Depending on the nature of the companion star, the most popular progenitor scenarios of SNe Ia fall into three
general categories: (i) a carbon-oxygen (CO) WD accretes matter from a non-degenerate companion, potentially a main-sequence (MS),
subgiant (SG), red giant (RG), or even helium (He) star, to trigger an explosion when 
its mass reaches the Chandrasekhar-mass (Ch-mass) limit. This is the single-degenerate (SD) scenario \citep{Whel73, Han04}, (ii) 
explosions are caused by the merger of two CO WDs. This is the double-degenerate (DD) scenario \citep{Iben84, Webb84},
(iii) the WD accretes material from a He-burning star or a He WD to lead to a Sub-Chandrasekhar-mass (Sub-Ch-mass) 
explosion when the He-shell accumulation reaches a critical value of about $0.02$--$0.2\,M_{\sun}$. This is the double-detonation scenario \citep{Livn90, Woos94}.

On the one hand, some narrow absorption signatures of circumstellar material (CSM) have been detected 
in some SNe Ia \citep{Pata07, Ster11, Dild12}. 
This has been suggested to be evidence that supports the SD scenario because such CSM is generally expected 
to exist around SN Ia progenitor as the result
of mass transfer from the companion star, as well as from WD winds.
On the other hand, there is some evidence in favour of the DD scenario, e.g., 
the non-detection of pre-explosion companion stars in normal SNe Ia \citep{Li11, Bloo12}, the 
absence of H/He features in the nebular spectra of SNe Ia \citep{Leon07, Lund13, Lund15, Shap13, Maguire15}, the lack of radio and X-ray emission around peak
brightness \citep{Bloo12, Brow12, Chom12,  Hore12, Marg14},
and the absence of a surviving companion star in SN Ia remnants \citep{Kerz09, Scha12}.

It has been widely proposed that very early-time observations of SNe Ia can be used to place strong constraints on their 
progenitor systems (e.g., \citealt{Kase10, Nuge11}; \citealt[][hereafter \citetalias{Rabinak12}]{Rabinak12}; \citealt{Brow12, Cao15, Olli15, Marion15}).
In particular, \citet[][]{Kase10} suggested that the strong excess emission can arise from the 
collision of SN ejecta with its companion star in the SD progenitor system, and it should be 
detectable in the 
ultraviolet (UV) and blue optical bands in the first few hours to days after the SN explosion under favorable viewing 
angles. Therefore, early UV observations are proposed as a direct way to 
test progenitor models of SNe Ia. The analysis of observed early light-curves has been carried out for many nearby SNe 
Ia to look for evidence of these shock 
emissions \citep{Hayd10, Brow12, Cao15, Marion15, Olli15}, but only a sublumious SN Ia (iPTF14atg, \citealt{Cao15}) and one normal event (SN~2012cg, \citealt{Marion15})
have been reported as likely detections of the signatures of the interaction between SN ejecta and a stellar companion.
Interestingly, \citet{Marion15} have shown that the analytical model of \citet{Kase10} with a $6.0\,M_{\odot}$ MS companion star 
could provide an explanation for the excess blue luminosity detected in SN~2012cg at about three days after the explosion. They therefore 
reported that this was the first evidence for emission from the interaction of SN ejecta with a stellar companion star for normal SNe Ia, supporting 
the SD scenario.

In this work, by comparing theoretical predictions from detailed binary evolution calculations and binary population synthesis (BPS) 
models for a variety of progenitor systems of SNe Ia (such as pre-explosion companion properties, early-time UV emissions, stripped
H/He mass, and so on) to different observations of SN~2012cg, we try to place some constraints on
its possible progenitor system.

\section{SN~2012\lowercase{cg}}
\label{sec:2012cg}

SN~2012cg was spectroscopically classified as a SN Ia \citep{Cenko12, Silverman12}. It was discovered on 
2012 May 17 (UT) in the nearby spiral galaxy NGC~4424 ($15.2\pm1.9$ Mpc, \citealt{Cortes06}) by the Lick Observatory 
Supernova Search \citep{Filippenko01}. By fitting early-time data of SN~2012cg, 
\citet{Silverman12} obtained that it reached the maximum $B$-band brightness on
2012 June $2.0\pm0.75$, and the peak magnitude in the $B$-band was $\rm{m_{B}} = 12.09\pm0.02$ mag, which corresponds 
to an absolute magnitude of $M_{\rm{B}} = -19.73\pm0.30$ mag. However, by measuring the rise and initial decline of the light curve
of SN 2012cg in the $B$, $V$, $R_{c}$, and $I_{c}$ filters, \citet{Munari13} found the $B$-band luminosity of SN~2012cg
peaked on 2012 June 4.5 with an absolute magnitude of $M_{\rm{B}}=-19.55\,\rm{mag}$ (after corrections for a reddening of $E(B-V)=0.18\,\rm{mag}$).
In addition, it was found that the light curve of SN~2012cg was slightly narrower than 
that of a typical normal SN Ia, namely SN~2011fe \citep{Silverman12}. They also estimated the explosion date of SN~2012cg to be May 15.7 (UT), 
which is consistent with the explosion time of MJD=56062.5 (May 15.5 UT) estimated by \citet{Marion15}.

\begin{figure}
  \begin{center}
    {\includegraphics[width=\columnwidth, angle=360]{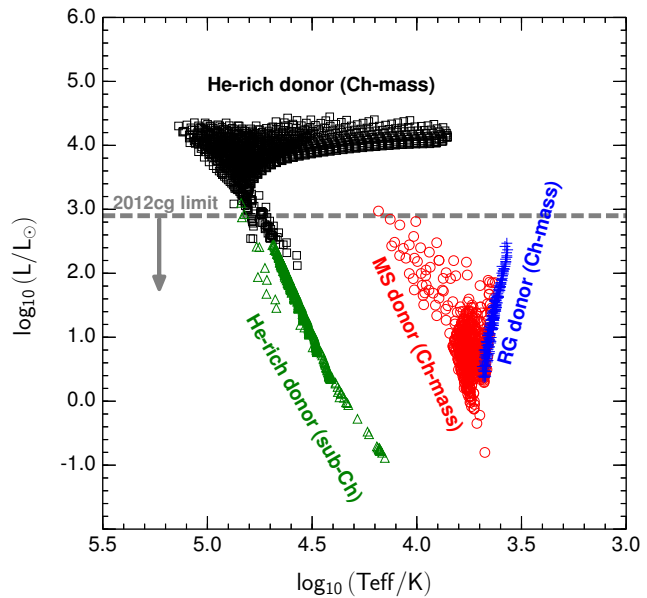}}
  \caption{H-R diagram of companion stars at the moment of SN explosion in different progenitor scenarios.  
           The red circles, blue crosses, and black squares
           correspond to the MS, RG, and He star donor Ch-mass scenario. The green triangles represent the sub-Ch-mass scenario. Here, only He-burning
           donor stars are considered in the Sub-Ch-mass scenario. The gray dashed line
           gives an upper-limit on the progenitor luminosity of SN~2012cg from $\textit{HST}$ pre-explosion observations \citep{Graur15}.}
\label{Fig:1}
  \end{center}
\end{figure}

\section{The pre-explosion progenitor}
\label{sec:pre-explosion}

WDs can only be observed directly in our own Milky Way and several very nearby galaxies because 
they would be faint. We therefore do not expect to generally detect the pre-explosion progenitors 
in the DD scenario. In the SD scenario, however, the companion stars are non-degenerate
which generally play a major role in determining the pre-explosion luminosities of progenitor systems. 
Therefore,  analyzing pre-explosion images at the SN position provides a direct way to put constraints on the nature 
of the progenitor companion star (e.g., \citealt{McCu14, Fole14}). However, no progenitors of normal SNe Ia have 
yet been directly observed, even for the relatively nearby events, SN~2011fe \citep{Li11} 
and SN~2014J \citep{Kell14}, although the probable progenitor system of a SN Iax SN~2012Z (i.e., SN~2012Z-S1) 
has been recently discovered \citep{McCu14}.

\begin{figure*}
  \begin{center}
    {\includegraphics[width=\columnwidth, angle=360]{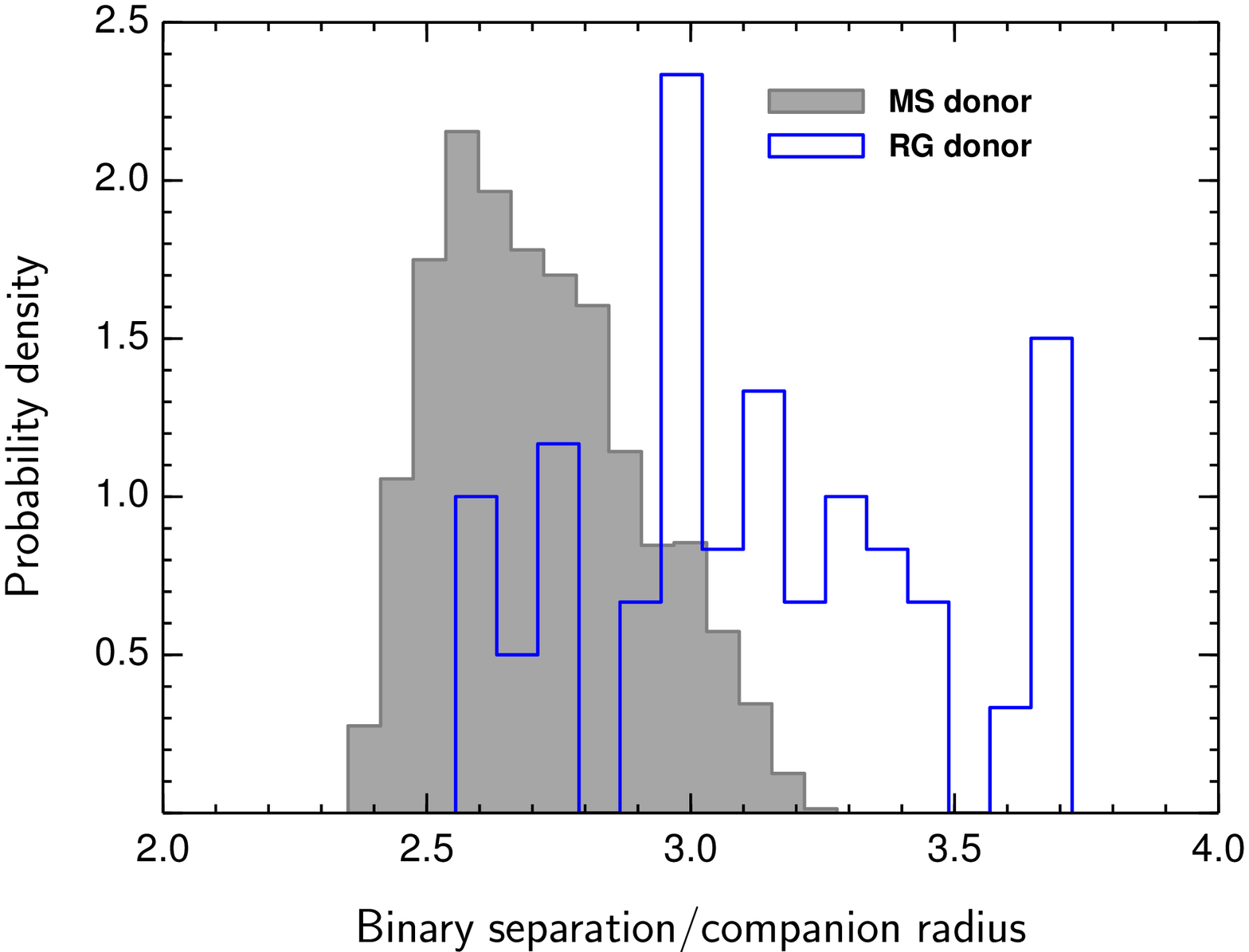}}
    \vspace{0.1in}
    {\includegraphics[width=\columnwidth, angle=360]{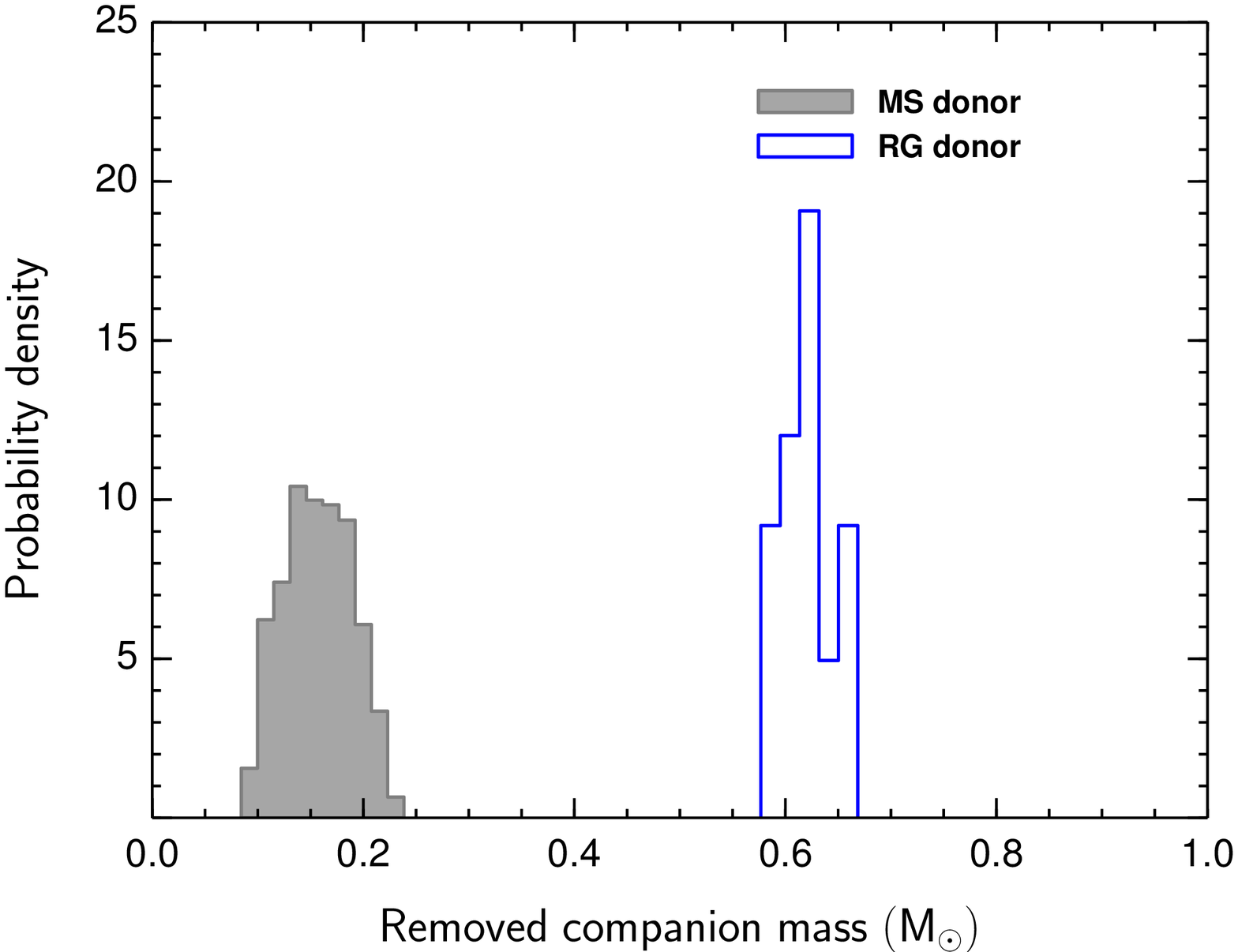}}
  \caption{\textbf{Left:} Distribution of the ratio of binary separation to companion radius ($a/R_{2}$) in the population 
                 synthesis calculations for the MS and RG donor Ch-mass scenario of SNe Ia (\citetalias{Liu15c}). 
                 \textbf{Right:} Distributions of the amount of removed H-rich material due to the SN impact. Here, the 
                 power-law relations between the total removed H-rich mass by the SN explosion and binary separation from three dimensional 
                 hydrodynamical simulations for the MS (Eq.~2 of \citealt{Liu12}) and RG (Eq.~4 of \citealt{Pan12}) 
                 donor scenario are used.
            }
\label{Fig:2}
  \end{center}
\end{figure*}

By analyzing pre-explosion archival WFPC2 images of NGC~4424 in F606W and F814W band from the $\textit{Hubble Space Telescope (HST}$),
\citet{Graur15} found no source  within a 2'' radius of the location of SN~2012cg down to limits of $\rm{M_{V}\approx-6.0}$ 
and $\rm{M_{V}\approx-5.4}$ mags, and thus excluded most supergiants as potential binary companions of the progenitor of SN~2012cg.
By taking into account Galactic and host-galaxy extinctions for these filters,  they place an upper-limit on the luminosity of pre-explosion 
progenitor of SN~2012cg of $<840\,L_{\sun}$.

In our previous study, we have predicted pre-explosion companion signatures of different SD progenitor scenarios 
by performing detailed binary evolution calculations with the {\sc STARS} code \citep{Liu15b}. Here, to place constraints on the possible progenitor of SN~2012cg, 
we directly use the predicted results in \citet{Liu15b} to compare with its pre-explosion $\textit{HST}$ observations in Fig.~\ref{Fig:1}.
As shown, most of the companion stars 
in the He-rich donor Ch-mass scenario are more luminous than the upper-limit constraint on the progenitor 
luminosity of SN~2012cg. Therefore, the He donor Ch-mass scenario can be 
excluded from the possible progenitor of SN~2012cg.

Here, we note that our previous calculations \citep{Liu15b} focused on sublumious SNe Iax, in which all WDs 
were assumed to trigger weak pure deflagration explosions (as proposed by \citealt{Fink14}) when they increase their 
masses to get close to the Ch-mass explosion limit. However, because the WD are treated as a point mass in our binary
evolution calculations, it is impossible to determine the exact ignition condition of the Ch-mass WD. No matter whether the Ch-mass CO WD undergoes
a delayed-detonation explosion that matches normal SNe Ia or a weak deflagration explosion of 
SNe Iax, the progenitor properties at the moment of SN explosion are the same in our binary evolution calculations. 
Therefore, to directly take the predictions from \citet{Liu15b} for the comparison of SN~2012cg in
Fig.~\ref{Fig:1} is reasonable.

\section{Searching for stripped hydrogen}
\label{sec:Hydrogen}

In the DD scenario, the donor star is another CO WD, which can naturally explain the lack
of H or He in the SN Ia spectra and the absence of a surviving companion star in SN remnants (SNRs). In the SD scenario, however,
the non-degenerate companion is a H- or He-rich star. After the SN explosion, the companion star 
is significantly hit and shocked by the SN ejecta, causing some H-rich or He-rich material to be removed from 
outer layers of the companion star, though the companion star survives the explosion (e.g., \citealt{Whee75, Mari00, Pakm08, Liu12, Liu13c, Pan12}).
Depending on the removed masses and the distances to observed SNe Ia, the signatures
of removed H or He may be detectable in late-time nebular spectra of some nearby SNe Ia.
Therefore, searching for $H_{\alpha}$ emission due to removed H-rich material in late-time spectra of SNe Ia 
is a way to identify the SD or DD progenitor scenario. Unfortunately, no such signature of swept-up H 
has been detected so far. This further places a stringent upper-limit 
on the removed H-mass of $0.001-0.058\,M_{\sun}$ \citep[e.g.,][]{Matt05, Leon07, Shap12,  Lund13, 
Lund15, Maguire15}.

Most interestingly, \citet{Maguire15} constrained the removed H-mass in SN~2012cg to be $\lesssim0.005-0.008\,M_{\sun}$, which 
is much less massive than the removed H-rich mass predicted from recent hydrodynamical simulations for the MS and RG donor 
scenario \citep[$\gtrsim\,0.1\,M_{\sun}$ with 
a typical velocity of $\lesssim1000\,\rm{km\,s^{-1}}$, e.g.,][]{Liu12, Liu13a, Pan12}. This poses a serious challenge to the MS/RG donor binary system as 
the progenitor of SN~2012cg. However, only several 
specific binary systems have been studied in these hydrodynamical simulations, and the parameter surveys in the simulations 
have shown that the amount of removed companion mass strongly decreases as the binary separation increases. 
Different binary systems evolve to different evolutionary stages and have different binary parameters when the
WD explodes as an SN Ia. Consequently, the companion radii and binary separations of systems are expected to differ significantly
from the MS or RG companion star models used in hydrodynamical impact simulations. Therefore, a comparison between the observational upper-limit 
on the removed H-mass for SN~2012cg and the predicted distribution of removed H-rich masses from binary population synthesis (BPS) 
calculations for the MS and RG donor scenario is required.

To predict the early UV luminosity distribution due to the interaction of SN ejecta with a stellar companion star, we have performed BPS 
calculations for the MS, RG and He donor Ch-mass scenario of SNe Ia \citep[][hereafter \citetalias{Liu15c}]{Liu15c}\footnote{
Here, we point out that the effect of the accretion disk of a WD is not considered in our binary evolution and population synthesis 
calculations because it is not clear whether a disk instability
can occur (see \citealt{Kato12}).}.
Taking the data from \citetalias{Liu15c}, the distribution of the ratio of binary separation ($a$) to the companion radius ($R_{2}$), $a/R_{2}$, at the moment of SN Ia explosion is 
shown in Fig.~\ref{Fig:2}. Based on the distribution of $a/R_{2}$, we further calculate the distribution of removed H-rich mass due to
ejecta impact in the MS and RG donor scenario by adopting the power-law relations between the removed H-rich mass and binary separation from 
past hydrodynamical impact simulations for the MS (Eq.~2 of \citealt{Liu12}) and RG (Eq.~4 of \citealt{Pan12}) donor scenario. 
As shown in Fig.~\ref{Fig:2}, the peaks of removed H-rich  mass in the MS and RG donor Ch-mass scenario are about $0.15\,M_{\sun}$ 
and $0.60\,M_{\sun}$ respectively. These peak values are larger than the observed upper-limit of SN~2012cg by a factor
of around 20--100. Therefore, non-detection of removed H in late-time nebular spectra of SN~2012cg indicates that the MS or RG donor 
binary system is unlikely to be its possible progenitor.

\begin{figure*}
  \begin{center}
    {\includegraphics[width=\columnwidth, angle=360]{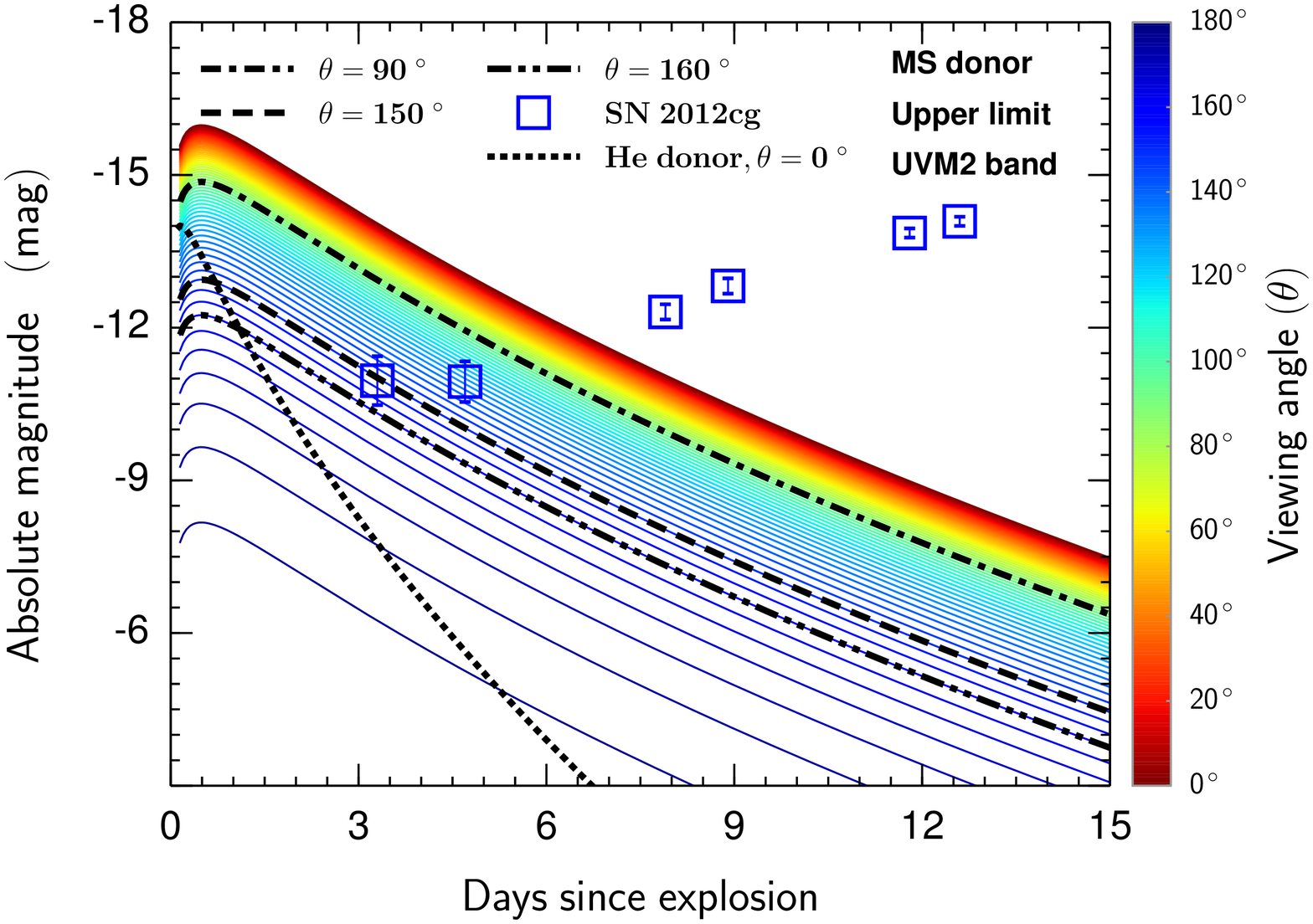}}
    \vspace{0.1in}
    {\includegraphics[width=\columnwidth, angle=360]{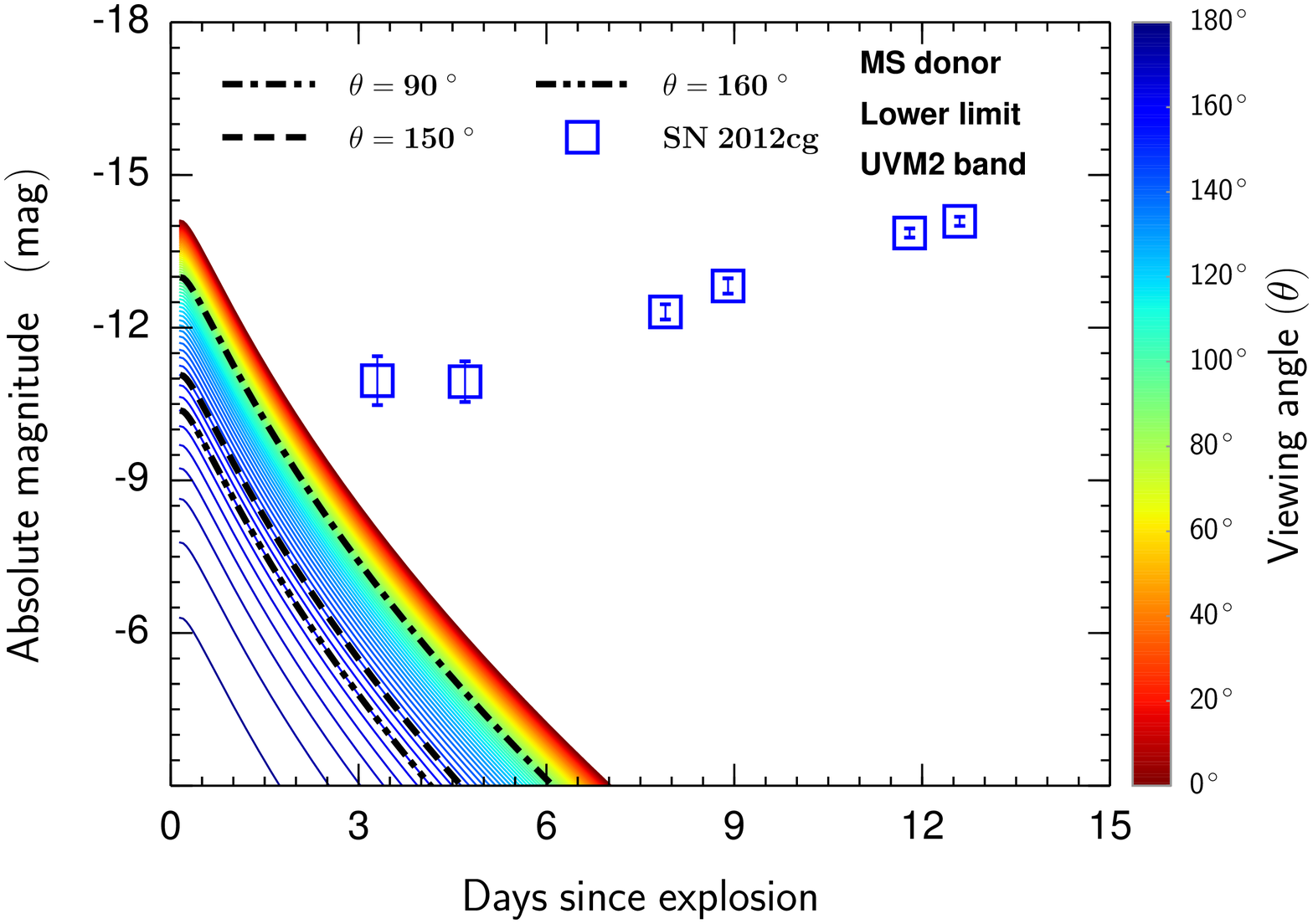}}
  \caption{Distributions of theoretical light curves predicted from the interaction between SN Ia ejecta and its MS companion star 
           in the $uvm2$ band of the $\textit{Swift}$ telescope. Here, the upper-limit and lower-limit (which corresponds to the upper 
           and lower boundary of a shaped region of the left-bottom panel in figure~2 of \citetalias{Liu15c}) luminosity evolution in the $uvm2$ band as 
           a function of viewing angle $\theta$ are shown in the left panel and right panel, respectively. Different colour means different viewing angle. The maximum UV flux (red color) 
           corresponds to a viewing angle of $\theta=0^{\circ}$ (i.e., looking down on the companion star).  For better visibility, the results for three particular viewing angles 
           are shown with black dashed, dash-dotted and dash-double-dotted lines. The upper-limit UV light curve for 
           a viewing angle $\theta=0^{\circ}$ in the He donor Ch-mass scenario is also plotted with a black dotted line in left panel.
           The early-time light curve of SN~2012cg in the $uvm2$ band from \citet{Marion15} is given with the square in blue for comparison.
           In our calculations, we assume an ejecta mass of $M_{\rm{ej}}=1.4\,M_{\rm{\sun}}$ and
           an explosion energy of $E_{\rm{ej}}=1.0\times10^{51}\,\rm{erg}$. 
            }
\label{Fig:3}
  \end{center}
\end{figure*}

\begin{figure*}
  \begin{center}
    {\includegraphics[width=\columnwidth, angle=360]{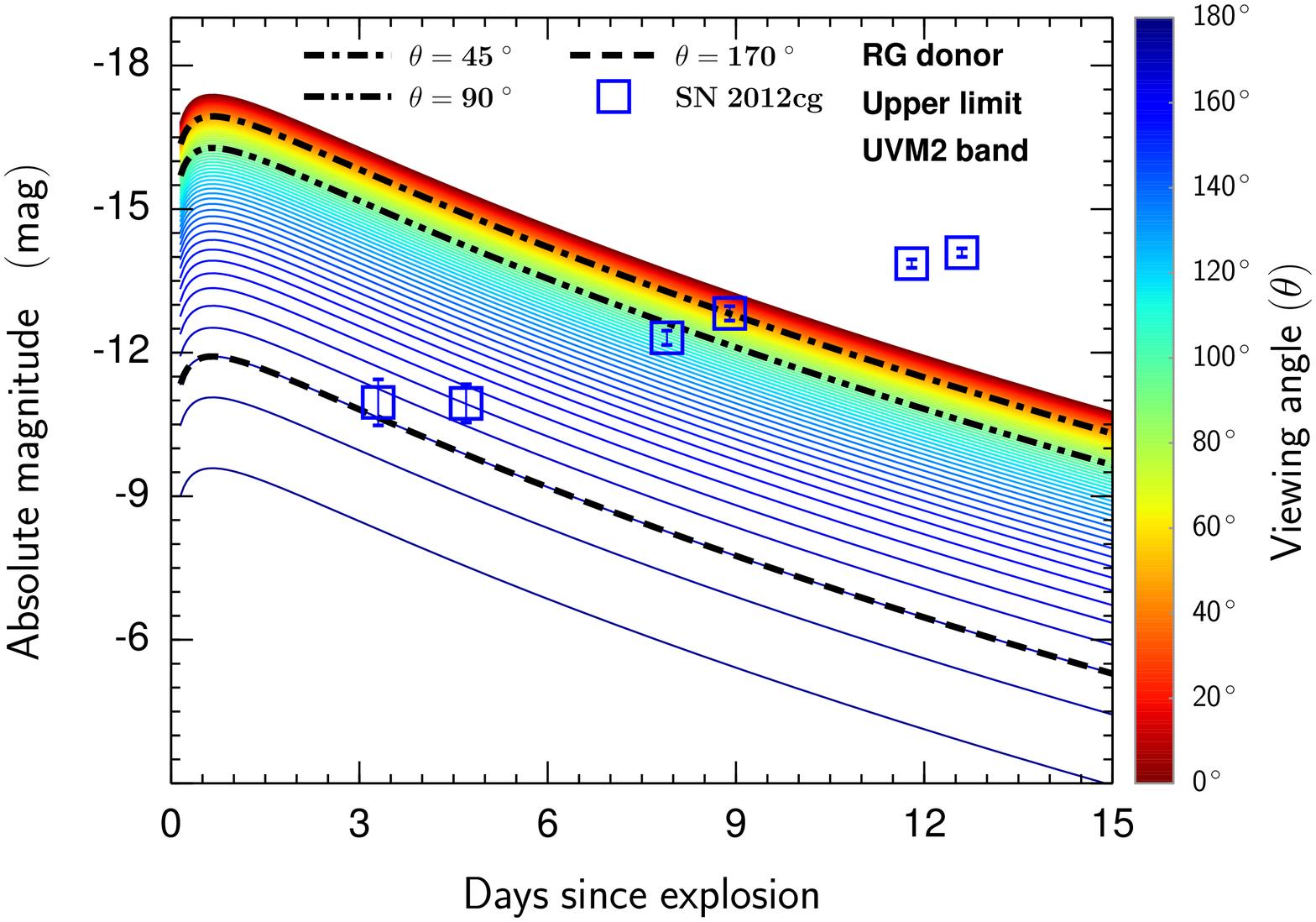}}
    \vspace{0.1in}
    {\includegraphics[width=\columnwidth, angle=360]{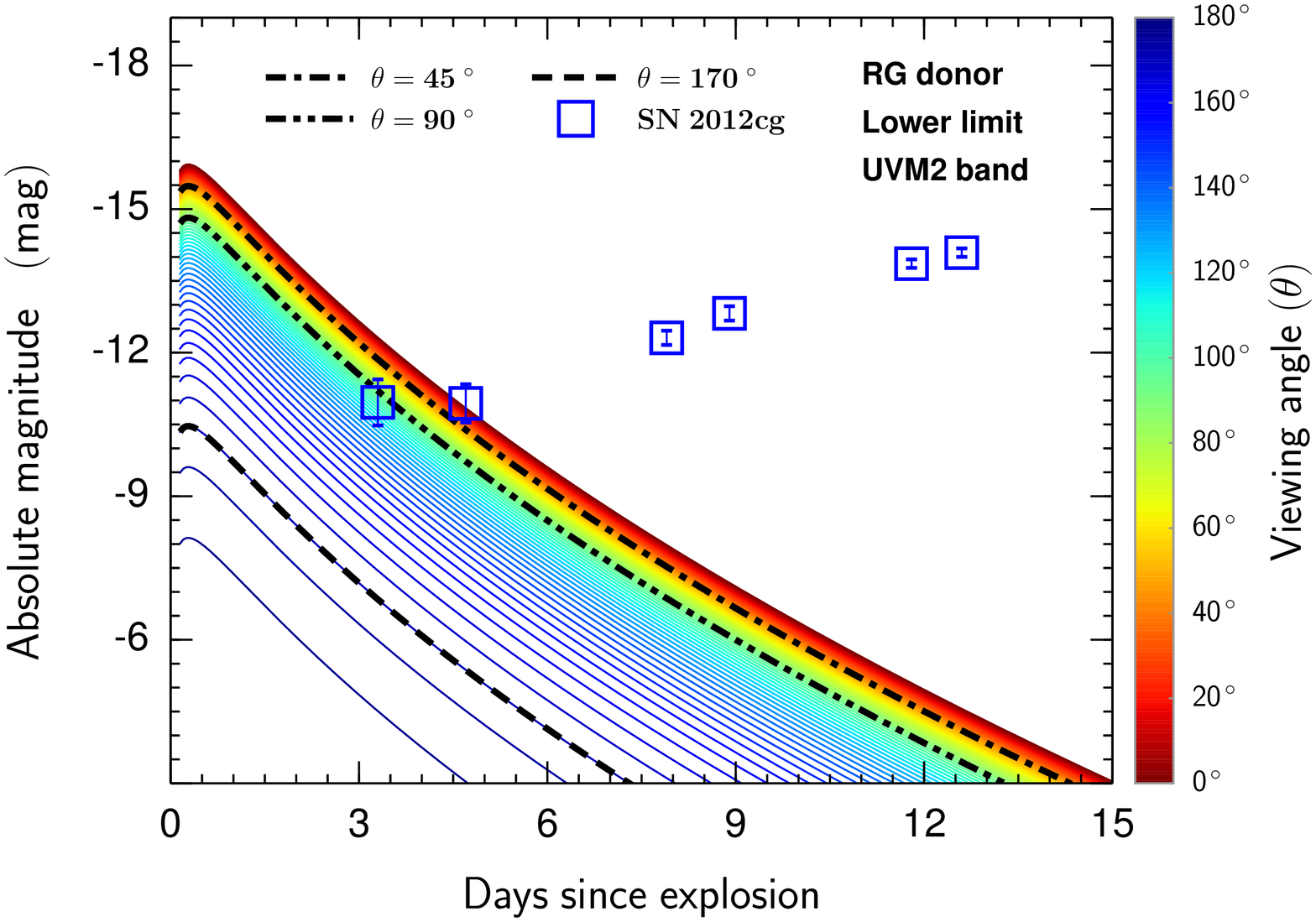}}
  \caption{As Fig.~\ref{Fig:3}, but for the RG donor Chandrasekhar-mass scenario.}
\label{Fig:4}
  \end{center}
\end{figure*}

Here, we point out that single fitting parameters of the power-law relations are used when we calculate the 
distributions of the removed H-rich masses by the SN explosion, which means that the effect of 
companion structures on the fitting parameters of the power-law relations are ignored. Even for the same progenitor scenario, it was found that fitting 
parameters are also dependent on the details of the structure of the companion star due to the history of mass transfer \citep{Liu12}. For example, for the MS donor scenario, 
the fitting parameters could be slightly different if the detailed structures of the MS companion star are somewhat different (see Table~3 of \citealt{Liu12}). 
However, we expect that the peak of removed H-rich mass in Fig.~\ref{Fig:2}  can only be shifted by about a factor of two
due to the uncertainties of fitting parameters among different MS (or RG) companion stars.

\section{Early-time excess luminosity}
\label{sec:uv}

\citet{Marion15} presented early-time photometry of SN~2012cg, showing that blue excess light is seen 
at about 16 days before maximum $B$--band brightness. They reported that this blue early-time excess is consistent 
with predicted early-time emission of shocked gas that was proposed by \citet{Kase10}, suggesting SN~2012cg was generated
from a binary system in the SD Ch-mass scenario. Here, we discuss the possibility of different origins of the early excess blue
luminosity of SN~2012cg.

\subsection{SN ejecta-companion interaction}

In \citetalias{Liu15c},  we presented the expected 
early UV luminosity distribution for different SD progenitor systems 
by applying the BPS results to the analytical model of \citet{Kase10}. 
Using the same method, the expected early luminosity distributions of 
shocked gas in the $\textit{uvm2}$ band for the MS, RG and He donor 
Ch-mass scenario are compared to an early-time light curve of SN~2012g in the same band (Figs.~\ref{Fig:3} and~\ref{Fig:4}) to place 
constraints on the possible progenitor system. Here, an ejecta mass of 
$M_{\rm{ej}}=1.4\,M_{\rm{\sun}}$ and an explosion energy of $E_{\rm{ej}}=1.0\times10^{51}\,\rm{erg}$ are used for our calculation. 
These typical values correspond to a mean expansion velocity of ejecta of $\approx10^{4}\,\rm{km\,s^{-1}}$. Also, the magnitudes 
are calculated in the AB magnitude system here.

\citet{Kase10} showed that the emission due to the interaction of SN ejecta with a companion star dominates the 
early-time (few days after explosion) light curves of SNe Ia within certain viewing angles. At later 
epochs, this shock flux becomes weak and the light curve is dominated by the flux of the SN itself. Also, 
this early-time excess emission should be brightest in the UV band and become subordinate at longer optical 
wavelengths although it can still cause a blue colour evolution in the optical light curve \citep{Kase10, Marion15}. 
Therefore, only predictions for the $uvm2$ filter (the $uvw1$ and $uvw2$ filters have an extended red tail) of the $\textit{Swift}$/UVOT is shown as an example.  
For a given binary progenitor system, the maximum excess emission should be observed when the viewing angle is $\theta=0^{\circ}$, i.e., the companion 
star lies directly along the line of sight between the observer and the SN  explosion. On the other hand, the excess emission 
detected for viewing angle $\theta=180^{\circ}$ should be negligible, corresponding to a geometry in which the SN Ia lies
directly in the line of sight between the observer and the companion star.

As shown in the left panel of Fig.~\ref{Fig:3}, the expected upper-limit luminosity at around three days after the explosion 
in the He donor Ch-mass scenario (black dotted line) is much lower than the observed value of SN~2012cg at a similar epoch. 
We therefore conclude that the early-time blue excess luminosity detected in SN~2012cg was unlikely to arise from 
the interaction of the SN shock with a companion star in the He donor Ch-mass scenario. 
Here, we would also like to mention that binary separations at the moment of SN explosions in the He star donor Sub-Ch-mass scenario 
are generally closer than those of the He star donor Ch-mass scenario, leading to early-time UV emission in this scenario being too 
weak to explain the early-time blue excess flux observed in SN~2012cg. Therefore, both the Sub-Ch-mass and Ch-mass
He star donor scenario could be ruled out if the early-time excess flux of SN~2012cg was indeed produced from the SN-companion interaction.

On the other hand, it is found that the expected upper-limit UV luminosity at three days after the explosion in the MS donor Ch-mass scenario decreases to 
a value below the observed excess luminosity of SN~2012cg as the viewing angle increases to a value of 
$\theta\approx150^{\circ}$ (Fig.~\ref{Fig:3}), suggesting that early-time emissions of the SN-companion interaction with a viewing 
angle of $\theta\approx0^{\circ}-150^{\circ}$ in this scenario would possibly provide an explanation for the early blue excess 
luminosity of SN~2012cg. Meanwhile, Fig.~\ref{Fig:4} shows that the expected lower-limit luminosity of shocked gas (right panel) 
in the RG donor scenario with a viewing angle larger than about $90^{\circ}$  is still higher than the observation of SN~2012cg. This indicates
that the viewing angle should be $90^{\circ}\lesssim\theta\lesssim170^{\circ}$ (left panel of 
Fig.~\ref{Fig:4}) if the early-time blue excess luminosity of SN~2012cg was indeed caused by the interaction of the SN shock with a RG donor 
star in the Ch-mass scenario.

\subsection{Cooling of shock-heated CSM}
\label{sec:sn-csm}

In the SD scenario, CSM is generally expected to be present around the progenitor 
due to pre-explosion mass loss in a binary system \citep{Pata07, Ster11, Dild12}. Also, some recent studies 
have suggested that a significant mass of CSM might be formed in the DD scenario: \vspace{-\topsep}
\begin{itemize} \itemsep1pt 
\item[(i)]$\textit{Tidal tail ejection}$. Prior to the coalescence of two WDs, \citet{Rask13} 
        found that about $\approx10^{-4}\,$--$\,10^{-2}\,M_{\sun}$ of material could be tidally ejected with a 
        typical velocity of about $2\,000\,\rm{km\,s^{-1}}$.  They further suggested that the interaction 
        of the SN ejecta with the ejected tails  could lead to detectable shock emission at radio, 
        optical/UV, and X-ray wavelenghts, and produce relatively broad
        \Nai\,D absorption features at late times.  
  \item[(ii)]$\textit{The double-detonation scenario.}$ \citet{Shen13} suggested that a total of $(3-6)\times10^{-5}\,M_{\sun}$
      of material can be ejected from a CO~WD~+~He~WD binary system to the local environment with a initial
      velocity of $\approx1500\,\rm{km\,s^{-1}}$. 
   \item[(iii)]$\textit{The violent merger scenario}$. It has been suggested that the 
         violent merger of a CO WD  with either the core of a giant star or another CO WD star
         might eject some material (about $0.1\,$--$\,0.5\,M_{\sun}$) and show some signs of CSM if the SN Ia 
         explodes soon after the common envelope phase \citep{Soke13, Ruit13}. 
   \item[(iv)]$\textit{Mass outflows during rapid accretion}$. 
         \citet{Dan11} found that some material ($\approx10^{-2}\,$--$\,10^{-3}\,M_{\sun}$) can be lost through the Lagrangian 
          point with a possible velocity of about $1000\,\rm{km\,s^{-1}}$ during the rapid accretion phase of two WDs.
   \item[(v)]$\textit{Disk winds}$. \citet{Ji13} found that a fraction of the 
          disk ($\approx10^{-3}\,M_{\sun}$) which forms before the SN 
          explosion if the WD-WD system fails to promptly detonate could be lost with a velocity of 
          around $2\,600\,\rm{km\,s^{-1}}$ due to a magnetically driven wind.
\end{itemize} 

Recently, \citet{Piro15} have shown that the presence of CSM around a progenitor system can lead to 
significant shock cooling emission during the first few days after the explosion, which can affect 
the early time rising light curve of a SN Ia. Depending on the degree of nickel mixing in the WD 
and the exact configuration of the extended material, this shock cooling emission can lead to early-time 
colour evolution similar to that caused by the ejecta interaction with a companion star \citep{Piro15}.
Therefore, the interaction of the SN shock with the CSM might provide an alternative explanation for the early blue colour 
seen in SN~2012cg \citep{Marion15}. Here, we adopt the analytical 
CSM-interaction model used in \citet{Cao15} to calculate the radius and mass of the CSM to examine the possibility 
of the early UV excess  of SN~2012cg having CSM-interaction as its origin.

As shown in \citet{Marion15}, the $\textit{Swift}$ measurements 
for SN~2012cg for the $B$- and $V$-band filters at about -16 days (MJD=56065.8) can fit those of SN~2011fe (or a $t^{2}$ model),  but 
measurements for SN~2012cg in the $uvm1$-, $uvm2$-, $uvw2$- and $U$-band 
are well above the observed values of SN~2011fe at the similar phase.
If SN~2011fe can be used as a template which only contains emissions from the SN 
itself (i.e., without any extra emission), then subtracting the flux of SN~2011fe from that of 
SN~2012cg at a similar epoch should give us the flux of the SN-CSM interaction. Subsequently, one could calculate the 
total UV luminosity ($L_{\rm{UV}}$) due to the SN-CSM interaction by summing the excess flux (i.e., the flux of SN~2012cg minus that of SN~2011fe at the 
similar epoch) over the $uvm1$-, $uvm2$-, $uvw2$- 
and $U$-band filters. With this method, we obtain an 
approximate excess UV luminosity of $L_{\rm{UV}}\approx 4.0\times10^{\,40}\,\rm{erg\,s^{-1}}$ at about 3.3 days 
after the explosion by using the total UV luminosity of SN~2012cg on MJD=56065.8 and subtracting that of SN~2011fe on 
MJD=55799.996. Here, an explosion date for SN~2012cg of MJD=56062.5 (i.e., May 15.5 UT, \citealt{Marion15}) is adopted, 
and the flux of SN~2011fe has been calibrated to the distance of SN~2012cg which is $15.2\,\rm{Mpc}$. Also, it should be kept in mind 
that we assume the actual explosion date of SN~2011fe was indeed on MJD=55796.696 \citep{Nuge11} through this study.   
Given this estimated excess UV luminosity, following \citet{Cao15} one can further calculate a spherical 
radius of the CSM of $R_{\rm{CSM}}\approx 1.0\times10^{16}\,\rm{cm}$ 
(or $R_{\rm{CSM}}\approx 9.0\times10^{15}\,\rm{cm}$), assuming a gas temperature 
of 15\,000\,K (or 20\,000\,K), leading to a CSM mass of $M_{\rm{CSM}}\approx1.2\,M_{\sun}$ (or $M_{\rm{CSM}}\approx0.83\,M_{\sun}$). 
Comparing all these values to theoretical predictions on properties of the CSM from different progenitor scenarios of SNe Ia,  a 
value of $M_{\rm{CSM}}\gtrsim0.8\,M_{\sun}$ at a distance of about $10^{16}\,\rm{cm}$ is larger than 
what most SD and DD scenarios predict \citep{Pata07, Ster11, Chom12, Rask13, Shen13, Soke13, Ruit13, Marg14}. 
This seems to indicate that the CSM-interaction as an origin of 
the observed early UV excess of SN~2012cg is quite unlikely. Here, instead of estimating the gas temperature by modelling 
the spectral energy distribution from the photometry and spectrum of SN~2012cg, we simply adopt a gas
temperature of 15\,000\,K (or 20\,000\,K) in our calculation. By modelling the near-UV to optical spectra 
of SN~2011fe with a Monte Carlo radiative transfer code, \citet{Mazzali14} obtained that the photospheric temperature 
of the photospheric blackbody at -16 days is about 10\,800\,K. In addition, adding a blackbody emission with 
a temperature of about 10\,000\,K to the spectra of SN~2011fe at about -16 days, \citet{Marion15} derived 
that the resulting continua can fit that of SN~2012cg at a 
comparable phase (see their figure~9). Therefore, adopting a gas temperature of 15\,000\,K--20\,000\,K should be 
safe for our calculations.

\begin{figure}
  \begin{center}
    {\includegraphics[width=\columnwidth, angle=360]{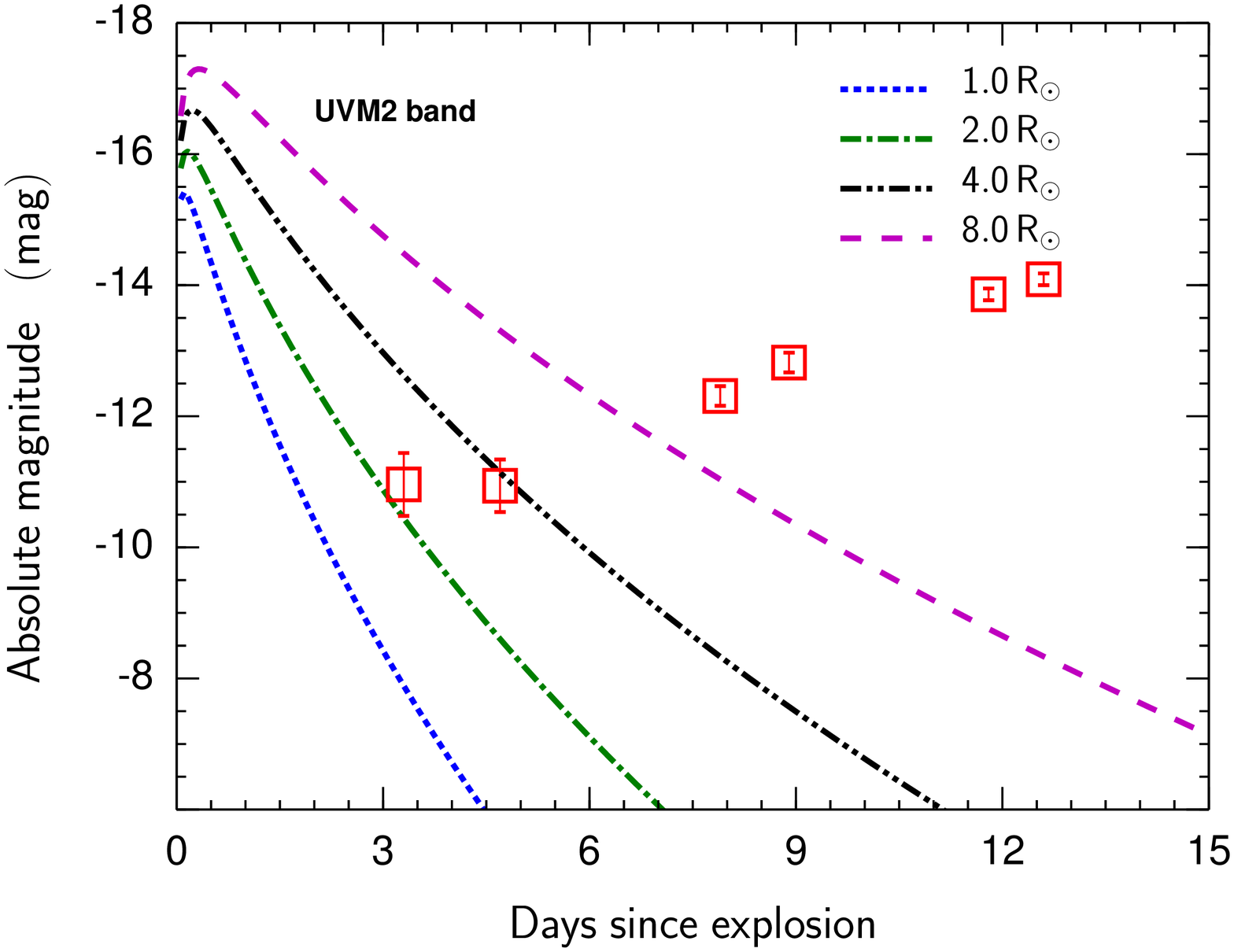}}
  \caption{Expected early-time light curve in $uvm2$-band from the diffusion
of radiation from the shock-heated ejecta by using the analytical model of \citetalias{Rabinak12}.
We assume an explosion energy of $10^{51}\rm{\,erg}$, an ejecta mass of $1.4\,M_{\sun}$, a form factor of 0.05, and an 
electron scattering opacity $0.2\,\rm{cm^2\,g^{-1}}$ \citep{Bloo12}. Different lines correspond to predictions 
with different initial progenitor radius. The observations of SN~2012cg are plotted with red square symbols. The absolute
magnitudes are calculated in the AB magnitude system.}
\label{Fig:5}
  \end{center}
\end{figure}

\begin{table*}
\centering
\caption{Summary of the constraints on the possible progenitor of SN~2012cg. }
 \begin{tabular}{@{}lllllllc@{}}
 \hline\hline
                          &\multicolumn{5}{c}{Possible progenitor of SN~2012cg?} &  \\
\hline
                          & MS donor  &  RG donor  & He star donor & He star donor & He WD donor & DD scenario & \\
Observational constraints & (Ch-mass) &  (Ch-mass) & (Ch-mass)     & (Sub-Ch-mass)      & (Sub-Ch-mass)    &             & References \\
\hline
Pre-explosion observations  & Yes	  &    Yes   &	No	 &  Yes	     & Yes      &  Yes       & [1]  \\
Early excess luminosity:     &          &       &       & 	     & &    &   \\
\ \ \ \  (i) Ejecta-star interaction    & Yes         &    Yes   & 	No       &  No	     & - &  -  & [2] \\
\ \ \ \  (ii) Ejecta-CSM  interaction   & Possible         &   Possible    & 	Possible       &  Possible   & Unlikely &  Possible  & [3]  \\
No stripped Hydrogen        & No	  &    No    &  Yes      &  Yes	     & Yes      &  Yes	     & [4] \\
Deep radio observations     & Possible    & Unlikely &  Possible &  Possible & Possible &  Possible  &  [5] \\
Late $^{57}\rm{Co}$ decay fitting        & Yes         &    Yes   &  Yes      &  No    & No    &  No     & [6]  \\ 
Narrow blueshifted absorption lines   & Possible         &   Possible   &  Possible &  Possible & Possible &  Possible  & [7]  \\

\hline
\end{tabular}

\medskip
\flushleft
\footnotesize
References: 
[1] \citet{Graur15, Li11, Liu15b}, 
[2] \citet{Kase10, Liu15c, Marion15}, 
[3] \citet{Cao15, Piro15}, 
[4] \citet{Matt05, Leon07, Shap13, Lund13, Maguire15, Mari00, Liu12, Liu13c, Pan12}, 
[5] \citet{Chom12, Chomiuk15}, 
[6] \citet{Graur15}, 
[7] \citet{Maguire13}. \\
\label{Tab:1}
\end{table*}

However, the above results are given with an explosion date for SN~2012cg of May 15.5 UT (MJD=56062.5, \citealt{Marion15}). If we assume
that the actual explosion date of SN~2012cg was one day earlier or later than May 15.5 UT, the date of MJD=56065.8 in SN~2012cg would correspond to 
4.3 or 2.3 days after the explosion, rather than the 3.3 days used above. We then adopt the above method to re-calculate the total excess UV luminosity due to the SN-CSM
interaction by 
using the flux of SN~2012cg on MJD=56065.8 and subtracting that of SN~2011fe at around 4.3 (MJD=55800.996, i.e., case A) or 
2.3 (MJD=55798.996, i.e., case B) days after the explosion, respectively.
As a result, for case A, we obtain an approximate  UV luminosity of $L_{\rm{UV}}\approx 1.0\times10^{\,40}\,\rm{erg\,s^{-1}}$, 
a radius of  $R_{\rm{CSM}}\approx 2.5\times10^{15}\,\rm{cm}$ (or $R_{\rm{CSM}}\approx 2.0\times10^{15}\,\rm{cm}$) 
and a CSM mass of $M_{\rm{CSM}}\approx0.07\,M_{\sun}$ (or $M_{\rm{CSM}}\approx0.05\,M_{\sun}$) 
for a gas temperature of 15\,000\,K (or 20\,000\,K), respectively. These values are 
consistent with predictions from most SD progenitor scenario \citep{Ster11, Marg14}.
As discussed by \citet{Cao15}, the SN shock 
has a typical velocity between $5\,000\,\rm{km\,s^{-1}}$ and $20\,000\,\rm{km\,s^{-1}}$. Hence, four days after the explosion, 
the SN shock was traveled to $R_{\rm{CSM}}\approx10^{15}\,\rm{cm}$, this is also consistent with our calculated results.
Comparing with the expected pre-explosion ejected mass from different DD scenarios presented above, the 
amount of ejected material from the He WD donor scenario  ($3-6\times10^{-5}\,M_{\sun}$, see \citealt{Shen13}) is  much less than our estimated CSM 
mass ($\approx0.05\,M_{\sun}$), which seems to suggest that the cooling of shock-heated CSM in the He WD donor scenario is unlikely to be the origin of the early blue excess 
flux in SN~2012cg. However, all other DD scenarios are possible, as they depend on the exact radius of the CSM (which is determined by the exact ejected velocity 
and the delay times between the mass ejected and the SN explosion). Unfortunately, both the ejected 
velocity and the delay time in the DD scenario are still poorly constrained.

In contrast, a calculation for case B gives a more massive CSM mass and a larger CSM radius, 
making the CSM-interaction as an origin of the early excess UV flux of SN~2012cg more unlikely. 
Therefore, whether the CSM interaction could provide a possible explanation for the early excess UV flux in 
SN~2012cg depends on the exact explosion date. \citet{Piro13} suggested that the SN may have a dark phase, which could
lead to the explosion date inferred from a simple light curve extrapolation being different from the actual explosion date.

\subsection{Cooling of the shock-heated WD}

After SN Ia explosion, radiation energy deposited in the shock-heated outer layers of the SN ejecta
immediately diffuses out and contributes to the SN brightness after the shock breakout \citepalias{Rabinak12}. The diffusion
of radiation from the shock-heated ejecta in the explosion is thought to be a contributor to the excess 
luminosity of SNe in addition to the luminosity generated by nickel heating. This shock breakout model has been 
used to put considerable constraints on the radius
of the progenitor star by detecting the early shock luminosity \citep{Nuge11, Bloo12}. 

In Fig.~\ref{Fig:5}, we compare the early-time observations of SN~2012cg in the $uvm2$-band to the predicted early-time light curves 
from the analytical model of \citetalias{Rabinak12} to examine the possibility 
that the early excess luminosity detected in SN~2012cg was produced from cooling of the shock-heated WD. 
Here, we assume an explosion energy of $10^{51}\rm{\,erg}$, an ejecta mass of $1.4\,M_{\sun}$, a form factor of 0.05, and an 
electron scattering opacity $0.2\,\rm{cm^2\,g^{-1}}$ \citep{Bloo12}. As shown in Fig.~\ref{Fig:5}, to account for
the early-time excess flux in SN~2012cg, a progenitor radius of $\approx2.0\,R_{\sun}$ is required. This is much larger
than the typical progenitor radius which is an isolated WD, in both the SD and DD scenarios, although some recent hydrodynamical studies 
suggested that the envelope of the exploding WD could spread out to a radius of $\approx0.1\,R_{\sun}$ in the violent merger
model \citep{Tanikawa15}. Therefore, one can conclude that the diffusion of radiation from the shock-heated ejecta in 
the shock breakout model of \citetalias{Rabinak12} is unlikely to be the contributor of the early excess luminosity seen in 
SN~2012cg.

\section{Late-Time Photometry}
\label{sec:late}

\citet{Seit09} suggested that significant contributions from the slow decay of $^{57}\rm{Co}$ and $^{55}\rm{Fe}$ can lead to 
a slow-down in the decline of the light curve of a SN Ia at very late times ($>$ 900 days after the explosion). Because different progenitor scenarios predict different 
amounts of $^{57}\rm{Co}$ and $^{55}\rm{Fe}$ during the explosion, different declines of late-time light curves should occur.
For instance, the delayed-detonation model of a Ch-mass WD produces more $^{57}\rm{Co}$ and $^{55}\rm{Fe}$ than those in the violent merger model of two WDs \citep{Roep12}.
Therefore, comparing the observed late-time light curves with the predictions has been suggested to be a potential way to constrain the progenitors of 
SNe Ia.

The late-time photometry of SN~2012cg has been followed using the $HST$ by \citet{Graur15}. They found its late-time light curve 
exhibits a similar behavior to that predicted by the decay of $^{57}\rm{Co}$ of a near Ch-mass explosion model,  
suggesting it favors the SD scenario. However, these authors also pointed out that the observed colour and light-curve shape 
at late times can also be explained by extra flux of a light echo. Therefore, future follow-up observations are needed to study whether 
its late-time photometry would still follow the slow decay of elements predicted from 
the Ch-mass WD explosion. This may help to exclude a light echo as an extra contributor to the late-time light curve 
of SN~2012cg and thus disfavor the DD scenario.

\section{Discussion}
\label{sec:discussions}

We summarise the constraints on the possible progenitor of SN~2012cg in Table~\ref{Tab:1}. For a comprehensive discussion, the results from other studies such as 
the deep radio observations of \citet{Chomiuk15}, the detection of blueshifted \Nai\,D \citep{Maguire13} and the late-time photometry observations \citep{Graur15} 
for SN~2012cg are also included in this table. As given in Table~\ref{Tab:1}, the Ch-mass He donor and RG donor scenario are
likely to be ruled out as the possible progenitor of SN~2012cg. Also, the Sub-Ch-mass scenario struggles to 
explain the excess blue luminosity seen in SN~2012cg. Meanwhile, current studies for the Sub-Ch-mass scenario found that this model still 
cannot reproduce the observational characteristic features of SNe Ia well  \citep{Krom10, Woos11}, although the details of 
this model still require future development and numerous complications remain to be solved in such a model. This scenario 
therefore can also be excluded.

Consequently, it seems that only the MS donor Ch-mass scenario and the merger of two CO WDs are likely to be 
the possible progenitor of SN~2012cg. However, both scenarios still have their own problems with explaining the observations.
The MS donor Ch-mass scenario predicts that a large amount of H-rich material (which is higher than the observational upper-limit of  
around $0.005\,M_{\sun}$ by a factor of 20) can be removed from outer layers of the companion star by the SN explosion, posing 
a serious challenge to the fact that no evidence of $H_{\alpha}$ emission in late-time spectra has been detected. However, as discussed by
\citet{Maguire15}, the observed flux at the position of $H_{\alpha}$ was converted to the mass with one-dimensional models by 
\citet{Matt05}, and \citet{Lund13} who performed spectral synthesis modelling based on the earliest two-dimensional impact simulation of \citet{Mari00}, 
which might lead to some uncertainties of the results. To place a more stringent constraint on the upper-limit mass for non-detection of swept-up H 
in late-time spectra of SNe Ia, future multi-dimensional radiative transfer modelling with ejecta structures obtained from 
updated three-dimensional impact simulations such as those of \citet{Liu12} and \citet{Pan12} is still needed.

Simultaneously, we have shown that the observed early excess blue flux in SN~2012cg could be explained by the SN-CSM interaction (Section~\ref{sec:sn-csm}).
However, this strongly depends on the exact explosion date. Also, both our estimations for the properties of the CSM in Section~\ref{sec:sn-csm} and calculations by \citet{Piro15} 
are based on an assumption of a spherical CSM structure. Whether the cooling of non-spherical CSM could still provide a possible explanation for the 
observed early blue flux of SN~2012cg is unclear. To place strict constraints on the properties of CSM and the observable signatures 
of the SN shock interacting with the CSM, detailed numerical hydrodynamical simulations 
for the interaction between the blast wave of the SN and CSM/ISM are required, coupled with different explosion models 
and the detailed pre-SN mass-loss history, i.e., different configurations of outflow material around progenitor
systems. In addition, \citep{Graur15} suggested that the decay of $^{57}\rm{Co}$ and $^{55}\rm{Fe}$ produced from current modelling 
for the two CO WDs merger model have difficulties in explaining the slow-down decline light curve of SN~2012cg at late times. 
However, they also pointed out that a similar behavior
of the late-time light curve of SN~2012cg can be explained by a light echo, which requires future follow-up observations
for further confirmation.

\begin{figure}
  \begin{center}
    {\includegraphics[width=\columnwidth, angle=360]{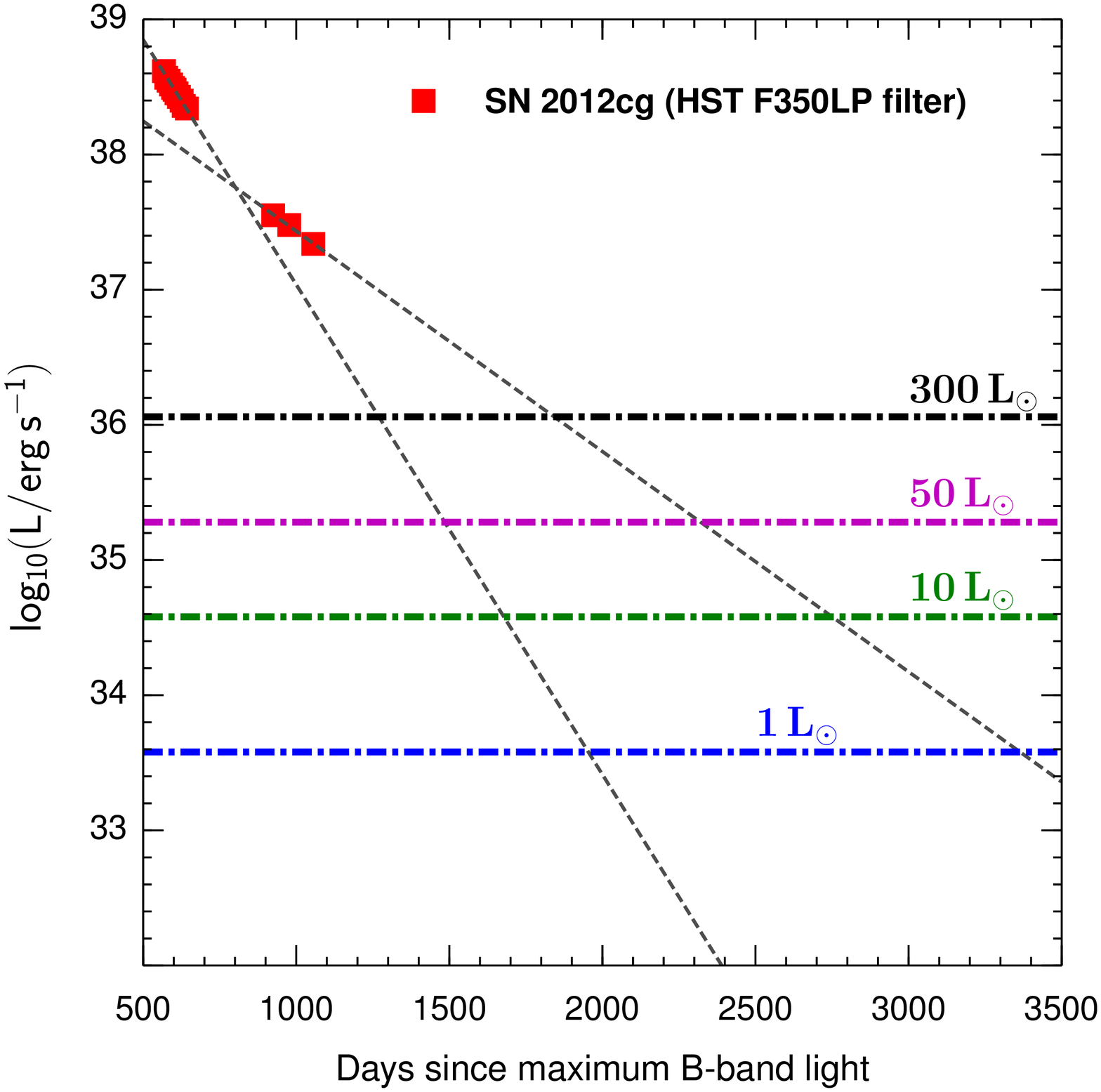}}
  \caption{Late-time light curve of SN~2012cg (square markers, \citealt{Graur15}) with linear late-time decline fit (gray dotted lines) and predicted 
          light curves (dash-dotted lines) of post-explosion companion stars 
          in the MS donor Ch-mass scenario \citep{Pan12, Shap13}. A MS companion star is expected to be overluminous (about 20--300\,$L_{\sun}$) 
          due to SN shock heating,  leading to that it would become more luminous than SN itself at very late times (around $2000\,\rm{days}$). For comparison, a star with a luminosity 
          of 1\,$L_{\sun}$ or 10\,$L_{\sun}$ is also plotted.}
\label{Fig:8}
  \end{center}
\end{figure}

\subsection{Future observations}

The companion stars in the MS 
donor Ch-mass scenario are expected to survive the explosion and to be significantly heated and shocked due to the interaction with 
SN ejecta, so that they would show some overluminous signatures (about 20--300\,$L_{\sun}$) during their long-term 
post-explosion evolution \citep{Pan12, Shap13}. As shown in Fig.~\ref{Fig:8}, the surviving MS star 
would be expected to become more luminous than the SN itself around $2\,000$ days after the maximum $B$-band brightness. 
Also, this surviving star is
predicted to move with a peculiar spatial velocity that is dominated by
its pre-explosion orbital velocity and possibly show some peculiar abundances signatures due to enrichment by heavy elements from 
SN ejecta \citep{Liu12, Pan12}. However, no surviving star is expected in the DD scenario. Therefore, 
searching for the post-explosion companion star in the SNR by future observations is a potential way to differentiate the SD and DD 
binary system as the progenitor of SN~2012cg.

\subsection{Spin-up/Spin-down model}

In the SD scenario, a WD accretes and retains companion matter that
carries angular momentum. As a consequence the WD spins with a short
period which leads to an increase of the critical explosion mass. If the critical
mass is higher than the actual mass of the WD, the SN explosion could only
occur after the WD increases its spin period with a specific spin-down
timescale \citep{Di11, Just11}. If the spin-down timescale is longer than about $10^{6}\,\rm{yrs}$, 
the CSM around the progenitor system could become diffuse and
reach a density similar to that of the ISM, leading to the lack of radio and X-ray emission 
which are absent in SN~2012cg \citep{Chomiuk15}. Also, the MS or RG companion star might shrink rapidly 
before the SN Ia explosion occurs by exhausting most of its H-rich 
envelope during a long spin-down ($\gtrsim10^{8}\,\rm{yrs}$) phase, explaining
the non-detection of a pre-explosion companion star \citep{Graur15} and the absence of swept-up H in the late-time spectra of SN~2012cg \citep{Maguire15}. However,
no (or a very weak) interaction signature from the SN-companion (because of a small companion) or SN-CSM (due to the low density CSM) interaction 
is predicted in this scenario, making it extremely difficult to  explain 
the early excess UV luminosity seen in SN~2012cg. Also, the exact spin-down timescale 
of the WD in this model is quite unknown \citep{Maoz14}.

\section{Conclusion and Summary}

\label{sec:conclusions}

In this study, using the results from binary evolution and population synthesis
calculations for the most likely progenitor scenarios of SNe Ia, we have tried to constrain
the possible progenitor of SN~2012cg. We find that the MS donor binary 
system in the SD scenario and the binary system consisting of two carbon-oxygen white dwarfs are more likely to 
be the progenitor of SN~2012cg. However, these two progenitor scenarios also predict theoretical results that 
are not consistent with the observations. For instance, a large amount of expected removed hydrogen mass in the MS donor scenario 
is in conflict with the absence of H lines in nebular spectra of SN~2012cg \citep{Maguire15}. On the other hand, whether the cooling of shock-heated circumstellar 
material in the double-degenerate scenario can provide a possible explanation for the observed early excess flux in SN~2012cg is still not 
well-constrained. Future observations and updated theoretical modelling will help to solve all these contradictions and thus 
place a strict constraint on the progenitor of SN~2012cg.     

\section*{Acknowledgments}
       We thank the anonymous referee for his/her valuable comments
       and suggestions that helped to improve the paper.
      ZWL thanks Yi Cao for his fruitful discussions. We would also like to thank 
      Carlo Abate and Takashi J. Moriya  for their helpful suggestions. This work is supported by the Alexander von Humboldt Foundation. 
      R.J.S. is the recipient of a Sofja Kovalevskaja Award from the Alexander von Humboldt Foundation.

\bibliographystyle{mn2e}

\bibliography{ref}

\begin{thebibliography}{76}
\expandafter\ifx\csname natexlab\endcsname\relax\def\natexlab#1{#1}\fi

\bibitem[{{Bloom} {et~al}\mbox{.}(2012){Bloom}, {Kasen}, {Shen}, {Nugent},
  {Butler}, {Graham}, {Howell}, {Kolb}, {Holmes}, {Haswell}, {Burwitz},
  {Rodriguez}, \& {Sullivan}}]{Bloo12}
{Bloom} J.~S. {et~al.}, 2012, \apjl, 744, L17

\bibitem[{{Brown} {et~al}\mbox{.}(2012){Brown}, {Dawson}, {de Pasquale},
  {Gronwall}, {Holland}, {Immler}, {Kuin}, {Mazzali}, {Milne}, {Oates}, \&
  {Siegel}}]{Brow12}
{Brown} P.~J. {et~al.}, 2012, \apj, 753, 22

\bibitem[{{Cao} {et~al}\mbox{.}(2015){Cao}, {Kulkarni}, {Howell}, {Gal-Yam},
  {Kasliwal}, {Valenti}, {Johansson}, {Amanullah}, {Goobar}, {Sollerman},
  {Taddia}, {Horesh}, {Sagiv}, {Cenko}, {Nugent}, {Arcavi}, {Surace},
  {Wo{\'z}niak}, {Moody}, {Rebbapragada}, {Bue}, \& {Gehrels}}]{Cao15}
{Cao} Y. {et~al.}, 2015, \nat, 521, 328

\bibitem[{{Cenko} {et~al}\mbox{.}(2012){Cenko}, {Filippenko}, {Silverman},
  {Gal-Yam}, {Pei}, {Nguyen}, {Carson}, \& {Barth}}]{Cenko12}
{Cenko} S.~B., {Filippenko} A.~V., {Silverman} J.~M., {Gal-Yam} A., {Pei} L.,
  {Nguyen} M., {Carson} D., {Barth} A.~J., 2012, Central Bureau Electronic
  Telegrams, 3111, 2

\bibitem[{{Chomiuk} {et~al}\mbox{.}(2015){Chomiuk}, {Soderberg}, {Chevalier},
  {Bruzewski}, {Foley}, {Parrent}, {Strader}, {Badenes}, {Fransson}, {Kamble},
  {Margutti}, {Rupen}, \& {Simon}}]{Chomiuk15}
{Chomiuk} L. {et~al.}, 2015, ArXiv:1510.07662

\bibitem[{{Chomiuk} {et~al}\mbox{.}(2012){Chomiuk}, {Soderberg}, {Moe},
  {Chevalier}, {Rupen}, {Badenes}, {Margutti}, {Fransson}, {Fong}, \&
  {Dittmann}}]{Chom12}
{Chomiuk} L. {et~al.}, 2012, \apj, 750, 164

\bibitem[{{Cort{\'e}s}, {Kenney} \& {Hardy}(2006){Cort{\'e}s}, {Kenney}, \&
  {Hardy}}]{Cortes06}
{Cort{\'e}s} J.~R., {Kenney} J.~D.~P., {Hardy} E., 2006, \aj, 131, 747

\bibitem[{{Dan} {et~al}\mbox{.}(2011){Dan}, {Rosswog}, {Guillochon}, \&
  {Ramirez-Ruiz}}]{Dan11}
{Dan} M., {Rosswog} S., {Guillochon} J., {Ramirez-Ruiz} E., 2011, \apj, 737, 89

\bibitem[{{Di Stefano}, {Voss} \& {Claeys}(2011){Di Stefano}, {Voss}, \&
  {Claeys}}]{Di11}
{Di Stefano} R., {Voss} R., {Claeys} J.~S.~W., 2011, \apjl, 738, L1+

\bibitem[{{Dilday} {et~al}\mbox{.}(2012){Dilday}, {Howell}, {Cenko},
  {Silverman}, {Nugent}, {Sullivan}, {Ben-Ami}, {Bildsten}, {Bolte}, {Endl},
  {Filippenko}, {Gnat}, {Horesh}, {Hsiao}, {Kasliwal}, {Kirkman}, {Maguire},
  {Marcy}, {Moore}, {Pan}, {Parrent}, {Podsiadlowski}, {Quimby}, {Sternberg},
  {Suzuki}, {Tytler}, {Xu}, {Bloom}, {Gal-Yam}, {Hook}, {Kulkarni}, {Law},
  {Ofek}, {Polishook}, \& {Poznanski}}]{Dild12}
{Dilday} B. {et~al.}, 2012, Science, 337, 942

\bibitem[{{Filippenko} {et~al}\mbox{.}(2001){Filippenko}, {Li}, {Treffers}, \&
  {Modjaz}}]{Filippenko01}
{Filippenko} A.~V., {Li} W.~D., {Treffers} R.~R., {Modjaz} M., 2001, in
  Astronomical Society of the Pacific Conference Series, Vol. 246, IAU Colloq.
  183: Small Telescope Astronomy on Global Scales, {Paczynski} B., {Chen}
  W.-P., {Lemme} C., eds., p. 121

\bibitem[{{Fink} {et~al}\mbox{.}(2014){Fink}, {Kromer}, {Seitenzahl},
  {Ciaraldi-Schoolmann}, {R{\"o}pke}, {Sim}, {Pakmor}, {Ruiter}, \&
  {Hillebrandt}}]{Fink14}
{Fink} M. {et~al.}, 2014, \mnras, 438, 1762

\bibitem[{{Foley} {et~al}\mbox{.}(2014){Foley}, {McCully}, {Jha}, {Bildsten},
  {Fong}, {Narayan}, {Rest}, \& {Stritzinger}}]{Fole14}
{Foley} R.~J., {McCully} C., {Jha} S.~W., {Bildsten} L., {Fong} W.-f.,
  {Narayan} G., {Rest} A., {Stritzinger} M.~D., 2014, \apj, 792, 29

\bibitem[{{Graur} {et~al}\mbox{.}(2016){Graur}, {Zurek}, {Shara}, {Riess},
  {Seitenzahl}, \& {Rest}}]{Graur15}
{Graur} O., {Zurek} D., {Shara} M.~M., {Riess} A.~G., {Seitenzahl} I.~R.,
  {Rest} A., 2016, \apj, 819, 31

\bibitem[{{Han} \& {Podsiadlowski}(2004)}]{Han04}
{Han} Z., {Podsiadlowski} P., 2004, \mnras, 350, 1301

\bibitem[{{Hayden} {et~al}\mbox{.}(2010){Hayden}, {Garnavich}, {Kasen},
  {Dilday}, {Frieman}, {Jha}, {Lampeitl}, {Nichol}, {Sako}, {Schneider},
  {Smith}, {Sollerman}, \& {Wheeler}}]{Hayd10}
{Hayden} B.~T. {et~al.}, 2010, \apj, 722, 1691

\bibitem[{{Hillebrandt} \& {Niemeyer}(2000)}]{Hill00}
{Hillebrandt} W., {Niemeyer} J.~C., 2000, \araa, 38, 191

\bibitem[{{Horesh} {et~al}\mbox{.}(2012){Horesh}, {Kulkarni}, {Fox},
  {Carpenter}, {Kasliwal}, {Ofek}, {Quimby}, {Gal-Yam}, {Cenko}, {de Bruyn},
  {Kamble}, {Wijers}, {van der Horst}, {Kouveliotou}, {Podsiadlowski},
  {Sullivan}, {Maguire}, {Howell}, {Nugent}, {Gehrels}, {Law}, {Poznanski}, \&
  {Shara}}]{Hore12}
{Horesh} A. {et~al.}, 2012, \apj, 746, 21

\bibitem[{{Hoyle} \& {Fowler}(1960)}]{Hoyl60}
{Hoyle} F., {Fowler} W.~A., 1960, \apj, 132, 565

\bibitem[{{Iben} \& {Tutukov}(1984)}]{Iben84}
{Iben}, Jr. I., {Tutukov} A.~V., 1984, \apj, 284, 719

\bibitem[{{Ji} {et~al}\mbox{.}(2013){Ji}, {Fisher}, {Garc{\'{\i}}a-Berro},
  {Tzeferacos}, {Jordan}, {Lee}, {Lor{\'e}n-Aguilar}, {Cremer}, \&
  {Behrends}}]{Ji13}
{Ji} S. {et~al.}, 2013, \apj, 773, 136

\bibitem[{{Justham}(2011)}]{Just11}
{Justham} S., 2011, \apjl, 730, L34+

\bibitem[{{Kasen}(2010)}]{Kase10}
{Kasen} D., 2010, \apj, 708, 1025

\bibitem[{{Kato} \& {Hachisu}(2012)}]{Kato12}
{Kato} M., {Hachisu} I., 2012, Bulletin of the Astronomical Society of India,
  40, 393

\bibitem[{{Kelly} {et~al}\mbox{.}(2014){Kelly}, {Fox}, {Filippenko}, {Cenko},
  {Prato}, {Schaefer}, {Shen}, {Zheng}, {Graham}, \& {Tucker}}]{Kell14}
{Kelly} P.~L. {et~al.}, 2014, \apj, 790, 3

\bibitem[{{Kerzendorf} {et~al}\mbox{.}(2009){Kerzendorf}, {Schmidt}, {Asplund},
  {Nomoto}, {Podsiadlowski}, {Frebel}, {Fesen}, \& {Yong}}]{Kerz09}
{Kerzendorf} W.~E., {Schmidt} B.~P., {Asplund} M., {Nomoto} K., {Podsiadlowski}
  P., {Frebel} A., {Fesen} R.~A., {Yong} D., 2009, \apj, 701, 1665

\bibitem[{{Kromer} {et~al}\mbox{.}(2010){Kromer}, {Sim}, {Fink}, {R{\"o}pke},
  {Seitenzahl}, \& {Hillebrandt}}]{Krom10}
{Kromer} M., {Sim} S.~A., {Fink} M., {R{\"o}pke} F.~K., {Seitenzahl} I.~R.,
  {Hillebrandt} W., 2010, \apj, 719, 1067

\bibitem[{{Leonard}(2007)}]{Leon07}
{Leonard} D.~C., 2007, \apj, 670, 1275

\bibitem[{{Li} {et~al}\mbox{.}(2011){Li}, {Bloom}, {Podsiadlowski}, {Miller},
  {Cenko}, {Jha}, {Sullivan}, {Howell}, {Nugent}, {Butler}, {Ofek}, {Kasliwal},
  {Richards}, {Stockton}, {Shih}, {Bildsten}, {Shara}, {Bibby}, {Filippenko},
  {Ganeshalingam}, {Silverman}, {Kulkarni}, {Law}, {Poznanski}, {Quimby},
  {McCully}, {Patel}, {Maguire}, \& {Shen}}]{Li11}
{Li} W. {et~al.}, 2011, \nat, 480, 348

\bibitem[{{Liu}, {Moriya} \& {Stancliffe}(2015){Liu}, {Moriya}, \&
  {Stancliffe}}]{Liu15c}
{Liu} Z.-W., {Moriya} T.~J., {Stancliffe} R.~J., 2015, \mnras, 454, 1192

\bibitem[{{Liu} {et~al}\mbox{.}(2013{\natexlab{a}}){Liu}, {Pakmor},
  {R{\"o}pke}, {Edelmann}, {Hillebrandt}, {Kerzendorf}, {Wang}, \&
  {Han}}]{Liu13a}
{Liu} Z.-W., {Pakmor} R., {R{\"o}pke} F.~K., {Edelmann} P., {Hillebrandt} W.,
  {Kerzendorf} W.~E., {Wang} B., {Han} Z.~W., 2013{\natexlab{a}}, \aap, 554,
  A109

\bibitem[{{Liu} {et~al}\mbox{.}(2012){Liu}, {Pakmor}, {R{\"o}pke}, {Edelmann},
  {Wang}, {Kromer}, {Hillebrandt}, \& {Han}}]{Liu12}
{Liu} Z.~W., {Pakmor} R., {R{\"o}pke} F.~K., {Edelmann} P., {Wang} B., {Kromer}
  M., {Hillebrandt} W., {Han} Z.~W., 2012, \aap, 548, A2

\bibitem[{{Liu} {et~al}\mbox{.}(2013{\natexlab{b}}){Liu}, {Pakmor},
  {Seitenzahl}, {Hillebrandt}, {Kromer}, {R{\"o}pke}, {Edelmann},
  {Taubenberger}, {Maeda}, {Wang}, \& {Han}}]{Liu13c}
{Liu} Z.-W. {et~al.}, 2013{\natexlab{b}}, \apj, 774, 37

\bibitem[{{Liu} {et~al}\mbox{.}(2015){Liu}, {Stancliffe}, {Abate}, \&
  {Wang}}]{Liu15b}
{Liu} Z.-W., {Stancliffe} R.~J., {Abate} C., {Wang} B., 2015, \apj, 808, 138

\bibitem[{{Livne}(1990)}]{Livn90}
{Livne} E., 1990, \apjl, 354, L53

\bibitem[{{Lundqvist} {et~al}\mbox{.}(2013){Lundqvist}, {Mattila}, {Sollerman},
  {Kozma}, {Baron}, {Cox}, {Fransson}, {Leibundgut}, \& {Spyromilio}}]{Lund13}
{Lundqvist} P. {et~al.}, 2013, \mnras, 435, 329

\bibitem[{{Lundqvist} {et~al}\mbox{.}(2015){Lundqvist}, {Nyholm}, {Taddia},
  {Sollerman}, {Johansson}, {Kozma}, {Lundqvist}, {Fransson}, {Garnavich},
  {Kromer}, {Shappee}, \& {Goobar}}]{Lund15}
{Lundqvist} P. {et~al.}, 2015, \aap, 577, A39

\bibitem[{{Maguire} {et~al}\mbox{.}(2013){Maguire}, {Sullivan}, {Patat},
  {Gal-Yam}, {Hook}, {Dhawan}, {Howell}, {Mazzali}, {Nugent}, {Pan},
  {Podsiadlowski}, {Simon}, {Sternberg}, {Valenti}, {Baltay}, {Bersier},
  {Blagorodnova}, {Chen}, {Ellman}, {Feindt}, {F{\"o}rster}, {Fraser},
  {Gonz{\'a}lez-Gait{\'a}n}, {Graham}, {Guti{\'e}rrez}, {Hachinger},
  {Hadjiyska}, {Inserra}, {Knapic}, {Laher}, {Leloudas}, {Margheim},
  {McKinnon}, {Molinaro}, {Morrell}, {Ofek}, {Rabinowitz}, {Rest}, {Sand},
  {Smareglia}, {Smartt}, {Taddia}, {Walker}, {Walton}, \& {Young}}]{Maguire13}
{Maguire} K. {et~al.}, 2013, \mnras, 436, 222

\bibitem[{{Maguire} {et~al}\mbox{.}(2016){Maguire}, {Taubenberger}, {Sullivan},
  \& {Mazzali}}]{Maguire15}
{Maguire} K., {Taubenberger} S., {Sullivan} M., {Mazzali} P.~A., 2016, \mnras,
  457, 3254

\bibitem[{{Maoz}, {Mannucci} \& {Nelemans}(2014){Maoz}, {Mannucci}, \&
  {Nelemans}}]{Maoz14}
{Maoz} D., {Mannucci} F., {Nelemans} G., 2014, \araa, 52, 107

\bibitem[{{Margutti} {et~al}\mbox{.}(2014){Margutti}, {Parrent}, {Kamble},
  {Soderberg}, {Foley}, {Milisavljevic}, {Drout}, \& {Kirshner}}]{Marg14}
{Margutti} R., {Parrent} J., {Kamble} A., {Soderberg} A.~M., {Foley} R.~J.,
  {Milisavljevic} D., {Drout} M.~R., {Kirshner} R., 2014, \apj, 790, 52

\bibitem[{{Marietta}, {Burrows} \& {Fryxell}(2000){Marietta}, {Burrows}, \&
  {Fryxell}}]{Mari00}
{Marietta} E., {Burrows} A., {Fryxell} B., 2000, \apjs, 128, 615

\bibitem[{Marion {et~al}\mbox{.}(2016)Marion, Brown, Vinkó, Silverman, Sand,
  Challis, Kirshner, Wheeler, Berlind, Brown, Calkins, Camacho, Dhungana,
  Foley, Friedman, Graham, Howell, Hsiao, Irwin, Jha, Kehoe, Macri, Maeda,
  Mandel, McCully, Pandya, Rines, Wilhelmy, \& Zheng}]{Marion15}
Marion G.~H. {et~al.}, 2016, \apj, 820, 92

\bibitem[{{Mattila} {et~al}\mbox{.}(2005){Mattila}, {Lundqvist}, {Sollerman},
  {Kozma}, {Baron}, {Fransson}, {Leibundgut}, \& {Nomoto}}]{Matt05}
{Mattila} S., {Lundqvist} P., {Sollerman} J., {Kozma} C., {Baron} E.,
  {Fransson} C., {Leibundgut} B., {Nomoto} K., 2005, \aap, 443, 649

\bibitem[{{Mazzali} {et~al}\mbox{.}(2014){Mazzali}, {Sullivan}, {Hachinger},
  {Ellis}, {Nugent}, {Howell}, {Gal-Yam}, {Maguire}, {Cooke}, {Thomas},
  {Nomoto}, \& {Walker}}]{Mazzali14}
{Mazzali} P.~A. {et~al.}, 2014, \mnras, 439, 1959

\bibitem[{{McCully} {et~al}\mbox{.}(2014){McCully}, {Jha}, {Foley}, {Bildsten},
  {Fong}, {Kirshner}, {Marion}, {Riess}, \& {Stritzinger}}]{McCu14}
{McCully} C. {et~al.}, 2014, \nat, 512, 54

\bibitem[{{Munari} {et~al}\mbox{.}(2013){Munari}, {Henden}, {Belligoli},
  {Castellani}, {Cherini}, {Righetti}, \& {Vagnozzi}}]{Munari13}
{Munari} U., {Henden} A., {Belligoli} R., {Castellani} F., {Cherini} G.,
  {Righetti} G.~L., {Vagnozzi} A., 2013, \na, 20, 30

\bibitem[{{Nomoto}(1982)}]{Nomo82}
{Nomoto} K., 1982, \apj, 253, 798

\bibitem[{{Nugent} {et~al}\mbox{.}(2011){Nugent}, {Sullivan}, {Cenko},
  {Thomas}, {Kasen}, {Howell}, {Bersier}, {Bloom}, {Kulkarni}, {Kandrashoff},
  {Filippenko}, {Silverman}, {Marcy}, {Howard}, {Isaacson}, {Maguire},
  {Suzuki}, {Tarlton}, {Pan}, {Bildsten}, {Fulton}, {Parrent}, {Sand},
  {Podsiadlowski}, {Bianco}, {Dilday}, {Graham}, {Lyman}, {James}, {Kasliwal},
  {Law}, {Quimby}, {Hook}, {Walker}, {Mazzali}, {Pian}, {Ofek}, {Gal-Yam}, \&
  {Poznanski}}]{Nuge11}
{Nugent} P.~E. {et~al.}, 2011, \nat, 480, 344

\bibitem[{{Olling} {et~al}\mbox{.}(2015){Olling}, {Mushotzky}, {Shaya}, {Rest},
  {Garnavich}, {Tucker}, {Kasen}, {Margheim}, \& {Filippenko}}]{Olli15}
{Olling} R.~P. {et~al.}, 2015, \nat, 521, 332

\bibitem[{{Pakmor} {et~al}\mbox{.}(2008){Pakmor}, {R{\"o}pke}, {Weiss}, \&
  {Hillebrandt}}]{Pakm08}
{Pakmor} R., {R{\"o}pke} F.~K., {Weiss} A., {Hillebrandt} W., 2008, \aap, 489,
  943

\bibitem[{{Pan}, {Ricker} \& {Taam}(2012){Pan}, {Ricker}, \& {Taam}}]{Pan12}
{Pan} K.-C., {Ricker} P.~M., {Taam} R.~E., 2012, \apj, 750, 151

\bibitem[{{Patat} {et~al}\mbox{.}(2007){Patat}, {Chandra}, {Chevalier},
  {Justham}, {Podsiadlowski}, {Wolf}, {Gal-Yam}, {Pasquini}, {Crawford},
  {Mazzali}, {Pauldrach}, {Nomoto}, {Benetti}, {Cappellaro}, {Elias-Rosa},
  {Hillebrandt}, {Leonard}, {Pastorello}, {Renzini}, {Sabbadin}, {Simon}, \&
  {Turatto}}]{Pata07}
{Patat} F. {et~al.}, 2007, Science, 317, 924

\bibitem[{{Perlmutter} {et~al}\mbox{.}(1999){Perlmutter}, {Aldering},
  {Goldhaber}, {Knop}, {Nugent}, {Castro}, {Deustua}, {Fabbro}, {Goobar},
  {Groom}, {Hook}, {Kim}, {Kim}, {Lee}, {Nunes}, {Pain}, {Pennypacker},
  {Quimby}, {Lidman}, {Ellis}, {Irwin}, {McMahon}, {Ruiz-Lapuente}, {Walton},
  {Schaefer}, {Boyle}, {Filippenko}, {Matheson}, {Fruchter}, {Panagia},
  {Newberg}, {Couch}, \& {The Supernova Cosmology Project}}]{Perl99}
{Perlmutter} S. {et~al.}, 1999, \apj, 517, 565

\bibitem[{{Piro} \& {Morozova}(2015)}]{Piro15}
{Piro} A.~L., {Morozova} V.~S., 2015, ArXiv:1512.03442

\bibitem[{{Piro} \& {Nakar}(2013)}]{Piro13}
{Piro} A.~L., {Nakar} E., 2013, \apj, 769, 67

\bibitem[{{Rabinak}, {Livne} \& {Waxman}(2012){Rabinak}, {Livne}, \&
  {Waxman}}]{Rabinak12}
{Rabinak} I., {Livne} E., {Waxman} E., 2012, \apj, 757, 35

\bibitem[{{Raskin} \& {Kasen}(2013)}]{Rask13}
{Raskin} C., {Kasen} D., 2013, \apj, 772, 1

\bibitem[{{Riess} {et~al}\mbox{.}(1998){Riess}, {Filippenko}, {Challis},
  {Clocchiatti}, {Diercks}, {Garnavich}, {Gilliland}, {Hogan}, {Jha},
  {Kirshner}, {Leibundgut}, {Phillips}, {Reiss}, {Schmidt}, {Schommer},
  {Smith}, {Spyromilio}, {Stubbs}, {Suntzeff}, \& {Tonry}}]{Ries98}
{Riess} A.~G. {et~al.}, 1998, \aj, 116, 1009

\bibitem[{{R{\"o}pke} {et~al}\mbox{.}(2012){R{\"o}pke}, {Kromer}, {Seitenzahl},
  {Pakmor}, {Sim}, {Taubenberger}, {Ciaraldi-Schoolmann}, {Hillebrandt},
  {Aldering}, {Antilogus}, {Baltay}, {Benitez-Herrera}, {Bongard}, {Buton},
  {Canto}, {Cellier-Holzem}, {Childress}, {Chotard}, {Copin}, {Fakhouri},
  {Fink}, {Fouchez}, {Gangler}, {Guy}, {Hachinger}, {Hsiao}, {Chen},
  {Kerschhaggl}, {Kowalski}, {Nugent}, {Paech}, {Pain}, {Pecontal}, {Pereira},
  {Perlmutter}, {Rabinowitz}, {Rigault}, {Runge}, {Saunders}, {Smadja},
  {Suzuki}, {Tao}, {Thomas}, {Tilquin}, \& {Wu}}]{Roep12}
{R{\"o}pke} F.~K. {et~al.}, 2012, \apjl, 750, L19

\bibitem[{{Ruiter} {et~al}\mbox{.}(2013){Ruiter}, {Sim}, {Pakmor}, {Kromer},
  {Seitenzahl}, {Belczynski}, {Fink}, {Herzog}, {Hillebrandt}, {R{\"o}pke}, \&
  {Taubenberger}}]{Ruit13}
{Ruiter} A.~J. {et~al.}, 2013, \mnras, 429, 1425

\bibitem[{{Schaefer} \& {Pagnotta}(2012)}]{Scha12}
{Schaefer} B.~E., {Pagnotta} A., 2012, \nat, 481, 164

\bibitem[{{Schmidt} {et~al}\mbox{.}(1998){Schmidt}, {Suntzeff}, {Phillips},
  {Schommer}, {Clocchiatti}, {Kirshner}, {Garnavich}, {Challis}, {Leibundgut},
  {Spyromilio}, {Riess}, {Filippenko}, {Hamuy}, {Smith}, {Hogan}, {Stubbs},
  {Diercks}, {Reiss}, {Gilliland}, {Tonry}, {Maza}, {Dressler}, {Walsh}, \&
  {Ciardullo}}]{Schm98}
{Schmidt} B.~P. {et~al.}, 1998, \apj, 507, 46

\bibitem[{{Seitenzahl}, {Taubenberger} \& {Sim}(2009){Seitenzahl},
  {Taubenberger}, \& {Sim}}]{Seit09}
{Seitenzahl} I.~R., {Taubenberger} S., {Sim} S.~A., 2009, \mnras, 400, 531

\bibitem[{{Shappee}, {Kochanek} \& {Stanek}(2013){Shappee}, {Kochanek}, \&
  {Stanek}}]{Shap13}
{Shappee} B.~J., {Kochanek} C.~S., {Stanek} K.~Z., 2013, \apj, 765, 150

\bibitem[{{Shappee} {et~al}\mbox{.}(2013){Shappee}, {Stanek}, {Pogge}, \&
  {Garnavich}}]{Shap12}
{Shappee} B.~J., {Stanek} K.~Z., {Pogge} R.~W., {Garnavich} P.~M., 2013, \apjl,
  762, L5

\bibitem[{{Shen}, {Guillochon} \& {Foley}(2013){Shen}, {Guillochon}, \&
  {Foley}}]{Shen13}
{Shen} K.~J., {Guillochon} J., {Foley} R.~J., 2013, \apjl, 770, L35

\bibitem[{{Silverman} {et~al}\mbox{.}(2012){Silverman}, {Ganeshalingam},
  {Cenko}, {Filippenko}, {Li}, {Barth}, {Carson}, {Childress}, {Clubb},
  {Cucchiara}, {Graham}, {Marion}, {Nguyen}, {Pei}, {Tucker}, {Vinko},
  {Wheeler}, \& {Worseck}}]{Silverman12}
{Silverman} J.~M. {et~al.}, 2012, \apjl, 756, L7

\bibitem[{{Soker} {et~al}\mbox{.}(2013){Soker}, {Kashi}, {Garc{\'{\i}}a-Berro},
  {Torres}, \& {Camacho}}]{Soke13}
{Soker} N., {Kashi} A., {Garc{\'{\i}}a-Berro} E., {Torres} S., {Camacho} J.,
  2013, \mnras, 431, 1541

\bibitem[{{Sternberg} {et~al}\mbox{.}(2011){Sternberg}, {Gal-Yam}, {Simon},
  {Leonard}, {Quimby}, {Phillips}, {Morrell}, {Thompson}, {Ivans}, {Marshall},
  {Filippenko}, {Marcy}, {Bloom}, {Patat}, {Foley}, {Yong}, {Penprase},
  {Beeler}, {Allende Prieto}, \& {Stringfellow}}]{Ster11}
{Sternberg} A. {et~al.}, 2011, Science, 333, 856

\bibitem[{{Tanikawa} {et~al}\mbox{.}(2015){Tanikawa}, {Nakasato}, {Sato},
  {Nomoto}, {Maeda}, \& {Hachisu}}]{Tanikawa15}
{Tanikawa} A., {Nakasato} N., {Sato} Y., {Nomoto} K., {Maeda} K., {Hachisu} I.,
  2015, \apj, 807, 40

\bibitem[{{Webbink}(1984)}]{Webb84}
{Webbink} R.~F., 1984, \apj, 277, 355

\bibitem[{{Wheeler}, {Lecar} \& {McKee}(1975){Wheeler}, {Lecar}, \&
  {McKee}}]{Whee75}
{Wheeler} J.~C., {Lecar} M., {McKee} C.~F., 1975, \apj, 200, 145

\bibitem[{{Whelan} \& {Iben}(1973)}]{Whel73}
{Whelan} J., {Iben}, Jr. I., 1973, \apj, 186, 1007

\bibitem[{{Woosley} \& {Kasen}(2011)}]{Woos11}
{Woosley} S.~E., {Kasen} D., 2011, \apj, 734, 38

\bibitem[{{Woosley} \& {Weaver}(1994)}]{Woos94}
{Woosley} S.~E., {Weaver} T.~A., 1994, \apj, 423, 371

\end{thebibliography}

\end{document}